# Atomic-resolution spectroscopic imaging of ensembles of nanocatalyst particles across the life of a fuel cell

Huolin L. Xin[1,*], Julia A. Mundy[2,*], Zhongyi Liu[3], Randi Cabezas[4], Robert Hovden[2], Lena Fitting Kourkoutis[2], Junliang Zhang[3], Nalini P. Subramanian[3], Rohit Makharia[3], Frederick T. Wagner[3], and David A. Muller[2,5]

1. Department of Physics, Cornell University, Ithaca, NY 14853, USA

2. School of Applied and Engineering Physics, Cornell University, Ithaca, NY 14853, USA

3. Electrochemical Energy Research Laboratory, General Motors, Honeoye Falls, NY 14472, USA

4. Electrical and Computer Engineering, Florida International University, Miami, FL 33174, USA

5. Kavli Institute at Cornell for Nanoscale Science, Ithaca, NY14853, USA

* These authors contributed equally to this work.

**The thousandfold increase in data-collection speed enabled by aberration-corrected optics allows us to overcome an electron microscopy paradox – how to obtain atomic-resolution chemical structure in individual nanoparticles, yet record a statistically significant sample from an inhomogeneous population. This allowed us to map hundreds of Pt-Co nanoparticles to show atomic-scale elemental distributions across different stages of the catalyst aging in a proton-exchange-membrane fuel cell, and relate Pt-shell thickness to treatment, particle size, surface orientation, and ordering.**

Bulk and reciprocal space measurements provide accurate ensemble averages of nanoparticle systems, yet in doing so lose the connections between microscopic degrees of freedom when integrating over the myriad of different particles in any representative sample. Out of necessity, nanoscale chemical imaging to date has relied on a handful of spectra collected from a few selected particles, as it often takes a few hours to record a spectral map of a single nanoparticle. However, nanoparticle systems—especially during electrocatalysis—are heterogeneous and have multiple competing processes running in parallel. Thus, identifying and quantifying dominant mechanisms requires statistics on scores to hundreds of particles in order to reliably connect the microstructure to the bulk properties. With the development of aberration-corrected scanning transmission electron microscopy (STEM)[1, 2] and efficient electron energy loss spectra (EELS) collection systems, elemental concentrations and chemical bonding information can now be collected roughly a thousand times faster than on a conventional microscope, allowing rapid and reliable 2-D mapping of chemical distributions at atomic resolution[3]. While much of the focus of aberration correction has been on producing increasingly small sub-Angstrom electron beams, here we instead stop at an atomic-sized beam and increase the usable beam current. This enabled us to collect over one million EELS spectra and map out the concentrations of all atomic species in hundreds of Pt-Co nanoparticles used as fuel cell electrocataysts. We can thus quantify and correlate internal ordering, facet termination and surface structure—nanoparticle by nanoparticle—to identify the dominant degradation chemistries that limit the catalyst's efficiency. These measurements that would have taken years to record, and thus be too slow to provide feedback in a rapidly evolving field, were now collected in sessions of a few hours to days.

Addressing the degradation mechanisms in Pt-Co nanoparticles is a key step in the development of proton-exchange-membrane fuel cells (PEMFCs), one of the most promising alternatives to fossil fuel-based internal combustion engines; PEMFCs produce electricity with water as the only by-product[4-6]. The commercialization of PEMFCs is hindered by the sluggish cathodic oxygen reduction reaction (ORR, $H_2+1/2O_2=H_2O$) despite the use of precious metal Pt catalysts. To overcome the cost barrier, Pt alloys (Pt-M, where M represents transition metals[7-12] or rare earths[13]) are generally used to reduce the Pt loading in the cathode. Some of the Pt-based alloy nanoparticles, such as the technologically relevant Pt-Co[5], show enhanced ORR activity at the beginning of the fuel cell lifetime; however, they are not yet stable enough to maintain their advantage over pure Pt nanocatalysts by the end of their targeted life times[5, 14-18]. This degradation during fuel cell operation is attributed to: dissolution of material from the catalyst particles leading to mass loss[19-21]; corrosion and collapse of the carbon support leading to particle agglomeration[19, 22]; a reduction in surface area for the remaining material as the average particle size increases; and a reduction in the specific activity from the remaining surface area as the particle composition and structure are altered[18, 23]. While the first two issues are better understood and common to both Pt and Pt-M alloys, our focus here is on the final two issues which are determined by the less-well understood microscopic underpinnings and interplay between particle growth, surface chemistry, and catalytic activity.

Understanding the coarsening of these nanocatalysts, which leads to the decline in specific activity during operations, is essential — for the Pt-Co nanoparticles studied here, the specific activity of the alloyed particles drops to that of pure Pt by the end of the DOE targeted lifetime (Supplementary Table S1 and S2). Two competing schools of thought have been proposed: Ostwald ripening by the transport of individual atoms[24, 25] or coalescence by movement and collisions of whole nanoparticles[26, 27]. To understand the relative importance of each mechanism for our system, we characterized the commercial $Pt_3Co$ electrocatalyst (see methods) before and after voltage cycling in an operational fuel cell to establish a microstructure/catalytic activity relationship throughout the fuel cell lifecycle. We find that not only are both Ostwald ripening and coalescence statistically significant contributors in our material, but that there is a synergistic interplay between the two, as Pt re-deposition is enhanced around coalesced particles compared to un-coalesced particles. A surprising consequence of the interplay is the suggestion that decreasing coalescence (by changing the support material for instance) might reduce the Ostwald ripening of the very-thick Pt shells correlated with the loss in specific activity.

In addition, we also investigated pre-cycled particles following a heat treatment to restore equilibrium structures and internal ordering to make contact with existing studies. This allowed us to compare our spatially-resolved experiments on nanoparticles to macroscopically-averaged single-crystal experiments and ab-initio theories of equilibrium surface structure in vacuum. It also illustrates the difference between the ideal structures predicted for thermodynamically stable surfaces in vacuum and the actual structure after exposure to liquids. We identified monolayer segregation of Pt on the {111} facets of annealed nanoparticles, consistent with previously reported vacuum-annealed bulk {111} crystal surfaces[11], however we also found no statistically significant Pt segregation was found on the {100} facets of the same nanoparticles in the ensemble, something that had not been possible to examine by bulk methods, although it is expected from Monte-Carlo simulations[11]. This provides a useful check of the precision and sensitivity of our real-space sampling.

We then perform an acid leaching of these equilibrium particles to distill the impact of the acidic environment of the cathode from the effects of voltage cycling. This allows us to distinguish the chemistry from the electrochemistry when directly comparing samples at the beginning and end of fuel cell lifetime. After exposure to acid, a three-monolayer Pt shell formed, surprisingly independent of crystal facet. The thickness of the shell indicates that the catalytic enhancement of acid leached Pt-Co nanoparticles over pure Pt cannot be due to a nearest neighbor electronic effect[28, 29].

It is important to note that considerable previous effort has been invested in determining the atomic distributions of Pt and Co within individual nanoparticles using annular dark-field (ADF) STEM[14, 23]. The ADF-STEM imaging technique with its strong mass-thickness sensitivity (a function of the number of atoms viewed in projection, and their atomic numbers) can rapidly screen many particles[30, 31]. While this works well for detecting heavy atoms in matrices of light atoms[31-33], the converse is often not the case. For example, while the emergence of a Pt-rich shell surrounding a $Pt_3Co$ core is expected as Co is soluble in acid whereas Pt is not[11], it is difficult to distinguish between Co atoms and Co vacancies near the surface against a Pt background solely with ADF images. Similarly, it would be difficult to distinguish between Co-rich regions and voids in a de-alloyed Pt/transition metal composite nanoparticle. The elastic-scattering cross section of Pt is 6-7 times that of Co and thus the electrons scattered by Pt dominate the signal. For $Pt_3Co$, only 2-6 of the projected 10-25 atoms in an atomic column of a 2-6 nm particle are Co; thus the contrast changes between stoichiometric $Pt_3Co$ and pure Pt are less than the scattering from a single Pt atom or vacancy on the surface. This ambiguity was resolved with EELS of the inelastically-scattered electrons. In particular, we used an EELS-optimized, 100kV, $5^{th}$-order aberration-corrected Nion STEM[3] with a 1-1.4 Å diameter electron beam to determine the Pt and Co elemental distributions independently and unambiguously from their unique core-level binding energies with sub-monolayer resolution (see methods). The key advances here are correcting probe aberrations to $5^{th}$ instead of $3^{rd}$ order, correcting key $3^{rd}$-order aberrations at the entrance to the spectrometer to increase the collection angle from 25 to 90 mrad, using a high-brightness cold field emission source and trading source size for increased beam current[3, 34].

While EELS spectroscopic imaging allows for a direct measurement of the Pt and Co distributions in a nanoparticle, the size of a spectroscopic data set with a 1340-channel spectrum at each pixel limited the number of pixels in each image. This forced us to image each nanoparticle sequentially, rather than simultaneously acquiring a large number of particles, as can be done for ADF images, which at 1 channel/pixel use 1340 times less memory. To avoid any sampling bias in selecting which particles to map in EELS, we also recorded a large-area ADF image to determine the particle size distribution for all 946 nanoparticles (Supplementary Fig S1) in the field of view. We then randomly selected a subset of roughly 1 in every 10 nanoparticles. Due to the larger variance in the tails of the particle-size distribution and almost uniform distribution of smaller particles (<5nm radius in ADF), we deliberately oversampled larger particles (>5 nm radius) in the 30,000 voltage-cycled sample to obtain better counting statistics relating to the coarsening mechanism. Our EELS measurements were then weighted by the population distribution (Supplementary Fig S2) such that we were not biased by sampling preferences; this ensures an accurate correspondence between our EELS results and the bulk electrochemical properties.

Figure 1 shows typical EELS spectroscopic images and outlines the consequences of the different processing pathways. Core-shell structures with pure Pt shells and Pt-Co alloy cores were observed in the pre-cycled "as-received" (Fig. 1a), heat plus acid-treated (Fig. 1d), and voltage-cycled (Fig. 1b) samples. A relatively uniform elemental distribution throughout the particle interior was observed from the heat-treated sample (Fig. 1c). As a resolution test in Fig. 2a, the atomic-resolution EELS maps show Pt and Co ordering in a heat-treated $L1_2$ Pt-Co nanoparticle, as does the accompanying ADF lattice image, although not all particles are ordered (Supplementary Fig S4). The presence of a surface Pt monolayer on some facets was observed in the heat-treated sample, as seen in Fig. 2b (atomic-resolution images) and Fig. 3 (statistical analysis). In Fig. 2b, a {111} terminated facet of a heat-treated Pt-Co nanoparticle shows the surface structure predicted by Stamenkovic and coworkers for the equilibrium structure in vacuum[11]; the surface atomic layer is Pt (bright in the ADF image) and second plane from the surface is Co rich (dark in the ADF image indicating Co enrichment or Pt vacancies). The accompanying EELS line profile resolves the ambiguity, showing the reduced intensity in the second plane is from preferential Co-segregation, not Pt-vacancies.

However, a unique strength of this technique is that it enables the mapping of the Pt shell for an ensemble of particles from which we can construct a statistically significant facet-by-facet analysis rather than relying on the single analysis in Fig. 2b. Figure 3 shows the effect of the acid leaching from a statistical comparison of the Pt shell thickness between the heat-treated and heat plus acid-treated samples (see Supplementary Figure S11 for the spectroscopic maps of the particles). Histograms are shown for {111}, {100}, and all nanoparticle facets (including a handful of what may be {110} facets and facets that could not be indexed because the particles were not on axis; here "{100}" includes both the {100} of the $L1_2$ ordered structure and the {200} of the disordered structure). For the equilibrium heat-treated sample (Fig. 3a), the Pt shell thickness on the {111} facets was 1.6 ± 0.5 Å (~1 monolayer) while that on the {100} was 0.6 ± 0.7 Å - a null result for segregation on {100} (two standard error of the mean, 95.4% confidence interval). This preferential segregation of one monolayer of Pt on the {111} planes agrees with studies of bulk surfaces as well as Monte Carlo simulations[11, 35] and ADF image analysis of selected $Pt_3Co$ nanoparticles[14, 23]. We are not aware of previous segregation studies on the {100} surface, although electrochemically the activities on the {100} and {111} $Pt_3Ni$ surfaces are very different[36]. However, we next probe the integrity of this segregated layer in an aggressive oxidizing acidic environment, which provides an accelerated proxy to mimic the chemical environment of the cathode.

Treatment of the thermally-equilibrated particles in an oxidizing acid medium forms a 5.7 ± 0.3 Å (~ 2.6 monolayers) thick Pt shell (Fig 3b). Unlike the equilibrium sample, the heat plus acid-treated sample exhibited a uniform Pt shell thickness on all facets, i.e., there was no statistically-significant difference in the Pt shell thickness on differently oriented facets. Thus it appears that the preference for a Pt skin on the {111} facets of the equilibrium particles does not survive in the cathode environment. This was unexpected because the density of surface sites is different for the different facets, suggesting leaching rates might be expected to be different as well. Possible explanations might be that either diffusion rates along different zone axes counteract this trend, or the leaching occurs through a disordered intermediate. The particles also take on a more rounded appearance as the {111} facets that dominated the thermally annealed particles shrink to become comparable in area to the other facets.

The uniform Pt-shell thickness was also independent of particle size (Supplementary Fig. S3). This statistical quantification of the Pt shell is particularly important as these Pt-Co particles still show a catalytic advantage over Pt particles, despite the fact that the top three surface layers are pure Pt. Our quantification of the shell thickness following exposure to the oxidizing acidic environment of the cathode suggests that the catalytic enhancement observed in these Pt-Co nanoparticles over pure Pt cannot be due to a nearest-neighbor, or even second-nearest-neighbor electronic effect as the outermost three atomic layers are pure Pt. Rather, as the Pt shell is coherently strained to the Pt-Co alloyed core, any electronic origin for the boon in catalytic activity is likely to be indirect, such as a change in bandwidth due to a change in lattice constant[8, 28, 29, 37-39]. The shell thickness can grow during electrochemical cycling[18, 23], however with increasing strain energy cost with shell thickness, strain enhancement cannot be sustained for arbitrarily-thick Pt shells and must relax beyond some critical thickness[40].

We now directly study the impact of electrochemical potential cycling by analyzing statistics from spectroscopic images of hundreds of nanocatalysts, before and after cycling. We also measured the electrochemical properties (i.e., hydrogen-adsorption-desorption area - HAD, mass activity – MA, and specific activity - SA) and particle size distributions prior to EELS spectroscopic imaging. The main electrochemical trends (Supplementary Table S1) are: i) Additional heat-treatment and acid-leaching on the as-received sample produced a drop in HAD and MA. This is perhaps because the manufacturer optimized the treatment of the as-received particles. ii) Additional heat-treatment and acid-leaching had insignificant influence on particle size distributions. iii) After 10,000 and 30,000 voltage cycles there were systematic reductions in the HAD, MA and SA, and increased particle growth. Figure 4 shows a collage of EELS spectroscopic images comparing the pre- and post-cycled samples. In Fig. 4a, the Pt and Co concentration maps show the pre-cycled, as-received particles have a thin relatively, uniform Pt-rich shell of 7.0±0.2 Å (~3.2 monolayers) surrounding the Pt-Co core. As expected, the shell in the as-received material is similar in thickness to that of the heat plus acid-treated sample (3.2±0.1 vs. 2.6±0.1 monolayers); unsurprisingly the specific activity was similar for these samples. In addition, the Pt-shell thickness appears to be independent of the particle size for the starting material (Fig. 5a). We also found that there was no correlation between the particle size and the presence of $L1_2$ ordering, nor does the ratio of ordered to disordered particles change significantly before and after voltage cycling (1.5±0.5 vs. 1.9±0.7, Supplementary Fig. S4).

In comparison to the starting, pre-cycled sample, the voltage-cycled sample has the following distinguishing features: i) a non-uniform distribution of Co from multiple Pt-Co cores within many of the particles (e.g., Fig. 4c); ii) much thicker Pt shells for some of the particles (e.g., Fig. 4d); iii) a strong correlation between Pt shell thickness and particle size (Fig. 5a); iv) the particles are more rounded, without clear or well-defined facets (a similar trend can also be seen between the heat treated and acid-leached particles - Supplementary Figure S11). For a bimetallic alloy such as Pt-Co, these varied spectroscopic maps provide important clues about the coarsening mechanism(s); without these chemical markers such mechanism would be non-trivial to deduce for pure Pt catalyst nanoparticles[24]. The key is that Co is soluble in acid and will be irreversibly leached from the Pt-Co nanoparticles[11]. In contrast, Pt can reversibly dissolve, re-deposit, and reduce via electrochemical processes during operation[14, 19]. (As shown in Table 1, nearly eight times as much Co is lost from the cathode region during the electrochemical cycling as Pt.) Thus, the presence of multiple Pt-Co cores in a single nanoparticle, e.g. Fig. 4c, suggests that coalescence was a contributing mechanism, while growth of particle size by adding a thick Pt shell that contains no Co, e.g. Fig. 4d, indicates Ostwald ripening. Continued leaching of Co from the interior of the particles would lead to a significant reduction in the Pt-Co core radius, which was not observed here. Skeletal structures formed by de-alloying instabilities can be recognized by the appearance of voids, and does not seem to be a major factor for the conditions studied here. In addition, a consequence of viewing thick three-dimensional samples in a two-dimensional projection creates the appearance that that there might be some aggregated particles prior to voltage cycling and that apparent "coalescence" could rather be Pt deposition onto neighboring particles. However, tomography (Supplementary Figure S5) reveals that these apparent aggregates in the pre-cycling projection image are actually well separated in 3-D. Once the possibility of accidental overlap has been excluded, the presence of multiple Co cores (such as those in Fig 4c) can be used as a signature of coalescence.

A scatter-plot comparison of the Pt-shell thickness – particle size correlation between the pre and post-cycled samples is presented in Fig 5a. Statistical summaries of the main trends are given in Fig 5b-d and Table 1. In sharp contrast to the small, uniform Pt-shells of the pre-cycled sample, by the end of the fuel cell's targeted lifetime, there is a strong dependence of shell thickness on particle size (Fig. 5a) with some of the large particles having the largest shell thickness. While the Co core remained essentially unchanged in the surviving particles ($3.0 \pm 0.2$ nm vs $2.9 \pm 0.2$ nm respectively), the most probable radius of the single-core particles increased by $0.6 \pm 0.1$ nm during cycling, which is entirely attributable to the $0.7 \pm 0.3$ nm increase in Pt shell thickness (Fig. 5d and Table 1) by Pt redeposition. Thus, we can deduce that the thick shells on single core particles, as shown in Fig 4d, are mostly the product of Ostwald ripening rather than continued Co leaching; continued Co leaching would have decreased the size of the Pt-Co core and not produced the observed increase in particle size for the single core particles. Instead much of the Co lost from the cathode (Table 1) seems to have come from the smaller particles that have dissolved away completely. This is again consistent with an Ostwald ripening mechanism where dissolved Pt is more likely to be redeposited on the larger particles, protecting their Co-containing cores, but exposing fresh Co to dissolution on the smaller particles.

As our spectroscopic images allow us to infer which of these mechanisms impacted each particle, we can statistically analyze the Ostwald-ripened particles separately from those that coalesced. In the initial distribution (Figure 4a), we were not able to locate any well-defined, multi-core particles in the survey region selected for EELS measurements.  In contrast, after voltage cycling, we observe that 48±13% of the larger particles (diameter > 6 nm) have multiple cores (Fig. 5b); these multi-core particles contribute 19% of the overall surface area  (Fig. 5c). (Here we plot the surface-area-weighted probability function as the surface is the catalytically active region.) The Pt shells on the coalesced particles grew by 1.6± 0.4 nm, double that of growth on the single core particles.  Combined with no significant change in the effective Co core radius from coalescence, the most probable particle radius of the coalesced particles grew by 1.4±0.3 nm (Table 1), again twice that of the single-core particles.  While particle coalescence, in and of itself, does not mandate a significant rise in shell thickness, nevertheless as shown in Fig. 5d, the Pt shells of the multi-core particles grew by double that of the single-core particles.  The dramatic increase in both particle size and Pt shell thickness for coalesced particles demonstrates an enhanced Pt re-deposition compared to single-core particles that have grown by Ostwald ripening alone.

With Co being more soluble than Pt, de-alloying of the nanoparticles is another possibility that should be considered.  In the simplest model, one might expect that leaching of Co from the surface will leave behind a vacancy-rich Pt shell that, given the diffusion rates for surface migration of Pt or for a vacancy to travel one to two unit cells at room temperature, will quickly collapse to a solid Pt shell and slow further diffusion.  However, when the etching rate of the Co is comparable to the surface diffusion rate of the Pt, a de-alloying instability can set in[41], with local clustering of Pt exposing interior Co, which in turn is leached away leaving a skeletal or "spongy" structure. Generally, the skeletal structure does not occur for particles smaller than the diffusion length (where Pt is just redistributed on the surface), or below a critical concentration of Co, and critical overpotential (that is a function of the Co concentration)[41]. We would not generally expect de-alloying for 20-30% Co concentrations at 1 V overpotential, however, some large "spongy" particles (as well as the more typical smaller convex particles) have been reported for similar starting material at a 1.2V overpotential, based on ADF images[23]. We have found a few large particles with very similar ADF images (Supplementary Figure S12). They account for roughly 4% of the larger (>10 nm diameter) particles and ~1% of the total particles after cycling, and a similar fraction of the as-received material, suggesting they were not produced by operation in the fuel cell, nor do they account for a significant fraction of particles.  Figure 4e-h (and Supplementary Figure S12) shows EELS maps from these spongy-appearing particles after voltage cycling.  While some pores do exists (Figures 4g,h), most of the dark areas in the ADF image that give the spongy appearance are Pt-Co alloys, while the lighter regions are either pure Pt or Pt-rich Pt-Co alloys.  This again illustrates the need for EELS in distinguishing between voids and possible light-element (such as Co or O) enrichment.

Exploiting the improved optics of aberration-corrected STEM to collect the elemental concentration map of an unprecedented number of Pt-Co electrocatalysts at different stages in the fuel cell lifetime allowed us to connect the microscopic surface structure, particle composition and the bulk electrochemical performance with statistical confidence.  The method should be useful in the study of most 3d Pt-M catalysts (except perhaps Pt-V where overlap of the V-L with the Pt-N edge may be problematic).  In both the starting material and in the acidic environment of the cathode, an ~ 3 monolayer Pt shell forms around a Pt-Co core.  This shell thickness was independent of terminating facet and notably the preferential segregation of a single monolayer of Pt onto the {111} surfaces of the equilibrium particles did not survive exposure to acid. The well-developed facets of the equilibrium structures also do not survive exposure to acid, with particles taking on more rounded structures. Both these observations may explain why the dramatic enhancement in specific activities reported for {111} over other surfaces for single crystals are so much smaller in nanoparticles [7, 36].

While these acid-leached particles show enhanced catalytic activity over pure Pt, it cannot be due to a nearest, or even second-nearest neighbor, electronic effect of the Co as the three outmost surface layers of the acid-leached Pt-Co particles are Pt. Rather, we find that the shell remains coherently strained to the Pt-Co core, suggesting that the observed boon in specific catalytic activity (compared to pure Pt) could be related to the spacing or arrangement of Pt atoms on the surface, and/or electronic effects such as the change in the d-band width due to a change in the lattice parameter[28], instead of to a direct electronic coupling to the Co atoms. It is worth noting that despite the solubility of Co, many small Pt-Co cores survive the voltage cycling; this is encouraging for strategies that reduce the Pt loadings required by straining a Pt shell to a non-Pt core.

We also compared the particles before and after voltage cycling to directly address the particle coarsening mechanisms leading to a loss in both catalytic surface area and specific activity: after 0.6-1.0V cycling, the specific activity of these Pt-Co particles approaches that of pure Pt. Under these operating conditions, we find well-rounded, rather than highly-percolated structures. We also find that both Ostwald ripening and coalescence are statistically significant contributors to the reduced surface area/$g_{Pt}$, with approximately half of the larger particles having undergone coalescence. Not only do the coalesced particles have a larger mean particle size than un-coalesced particles, but surprisingly 2.9±0.9 times more Pt (by volume) re-deposits on them, likely due to Pt re-depositing in a way to minimize the surface curvature of the more complex shape resulting from coalescence.  This thick Pt shell formed on the coalesced particles, and some of the others, likely explains the loss of a catalytic advantage over pure Pt.  It also demonstrates the complex interplay between the two mechanisms, suggesting that decreasing coalescence (by controlling interactions between support and nanoparticles) would likely also bring a decrease in Pt re-deposition and the resulting loss in catalytic activity.  The tendency of Pt re-deposition to form rounded particles may pose a serious durability challenge for strategies to improve the catalytic activity by engineering the nanoparticle shape, though the structural evolution may differ for particles that start with larger initial faceting or cycle through defined potential ranges.

The ability to record statistically meaningful distributions of microscopic information on inhomogeneous nanoparticles ensembles should prove useful not only for catalysis research, but also for a wider variety of nanoparticle systems where competing processes could be proceeding in parallel.

**Acknowledgements**

The authors acknowledge discussions with F. J. DiSalvo and H. D. Abruna and early TEM by A. E. Megrant. The RDE work at GM was performed by P. Gregorius. The work at Cornell (H. L. X., J. A. M, D. A. M) was supported as part of the Energy Materials Center at Cornell (EMC2), an Energy Frontier Research Center funded by the U. S. Department of Energy, Office of Science, Office of Basic Energy Sciences under Award Number DE-SC0001086. R. C. was supported by the Nanoscale Science and Engineering Initiative of the National Science Foundation under NSF Award EEC-0117770,064654. J.A.M. acknowledges support from an NDSEG fellowship. Facilities support by the Cornell Center for Materials Research (NSF DMR-0520404 and IMR-0417392) and NYSTAR.

**Author Contributions**

H.L.X. and J.A.M. contributed equally to this work. Z.Y.L. and D.A.M. designed the experiments. Electron microscopy, spectroscopy, tomography and data analysis were carried out by H.L.X., J.A.M., R.C., R.H., L.F.K. and D.A.M.; electrochemistry and fuel cell testing by J.Z., N.P.S. and Z.Y.L. All authors discussed the results and implications at all stages. D.A.M., Z.Y.L., J.A.M, and H.L.X. wrote the paper.

**Correspondence** and requests for materials should be addressed to D.A.M. (david.a.muller@cornell.edu) and Z.Y.L. (vic.liu@gm.com).

**Tables and Table captions in the text**

|  | As-received Pt-Co Particles | After 30k Cycling | 30k Cycled: Multi-core | 30k Cycled: Single-core |
|---|---|---|---|---|
| Particle Radius (nm) | 3.7±0.1 | 4.5±0.1 | 5.1±0.16 | 4.3±0.1 |
| Pt-Co Core Radius (nm) | 3.0±0.2 | 2.9±0.1 | 2.8±0.3 | 2.9±0.2 |
| Pt-Shell Radius (nm) | 0.7±0.2 | 1.6±0.14 | 2.3±0.3 | 1.4±0.2 |
| % of Pt Remaining in Electrode | 95±5% | 90±5% |  |  |
| % of Co Remaining in Electrode | 89±5% | 60±5% |  |  |

**Table 1**. Measured nanoparticle core/shell radii, weighted by particle volume to reflect the change in Co content before and after cycling, for the Pt-Co cores, the surrounding Pt shell and the entire particles comprising both Pt-Co core and the surrounding Pt shell. Also shown are bulk fractions of Pt and Co present in the electrode (with remainder in the membrane) determined by electron probe microanalysis before and after voltage cycling. The membrane electrode assembly (MEA) before cycling is listed in the as-received column, although the measurement here is on the actual MEA and not just the powder. Each is normalized separately to the total fraction of that element present in electrode and MEA, i.e. after voltage cycling 40% of the Co has migrated out of the electrode and into the membrane.

**Figure captions for figures in the text**

**Figure 1.** Typical EELS spectroscopic images of Pt-Co nanoparticles for the chemical and electrochemical processes studied here. The relative Pt concentration is shown in red and the Co concentration in green, so yellow indicates a Pt-Co alloy. The stripes in the Pt image of (**a**) are {111} lattice planes viewed at a slight mistilt. (**a**) as-received, showing 0.6 nm Co-free shell, (**b**) after 30k voltage-cycles, showing a coalescence of 2 smaller Pt-Co nanoparticles surrounded by a Co-free shell ~2-3 nm thick, (**c**) after heat-treatment of the as-received material, leading to a uniform Co distribution and no Co-free shell and (**d**) after acid-leaching the heat-treated sample, where a Co-free shell similar to that of the starting material (**a**) has returned. The scale bar for each image is 5 nm.

**Figure 2.** Atomic-resolution spectroscopic images of Pt-Co nanoparticles from the heat-treated sample. **a**, a $L1_2$ ordered Pt-Co nanoparticle with (001) planes oriented parallel to the electron beam. The Pt and Co maps show the alternating planes of the Pt and Pt/Co as shown in the model structure (Pt - red, Co - green). **b**, a {111}-facet terminated Pt-Co nanoparticle, and the Pt/Co concentration profile across the facet showing a strong de-alloying effect at the last two planes, with the Co segregating to one layer below the surface, and Pt segregating to outermost plane. The Pt and Co signals are scaled to their nominal bulk compositions at x=0.

**Figure 3.** Histograms of the Pt shell thickness, measured from different facets of individual nanoparticles in (**a**) the heat-treated sample and (**b**) the acid-leached sample. The left and right boundary of the colored shaded area marks the two standard error of the mean (95.4% confidence interval) and the right boundary of the gray shaded area marks the average shell thickness. The overlap of the red and gray areas for the {100} facets of the heat-treated sample shows that there is no statistically-significant shell thickness there, whereas there is a $1.6 \pm 0.5$ Å (~1 monolayer) Pt shell on {111} heat-treated facets. In contrast, after acid treatment (**b**) the shell thickness has grown to $5.7 \pm 0.3$ Å (~ 2.5 monolayers) and is independent of facet orientation. For the heat-treated sample, of the 94 facets measured, 21 could be indexed as {111}, and 15 as {100} while for the acid-leached material, of the 63 facets, 15 were identified as {111} and 12 as {100}.

**Figure 4.** Collage of EELS spectroscopic images from the (**a**) as-received and (**b**) the voltage cycled sample. As shown in the color bars, the relative Pt concentration, normalized in each separate image, is plotted in red and the Co concentration ranging from green to turquoise. Particles (**c**-**h**) from the cycled sample are discussed in more detail in the text. The selection of images shown is meant to display the diversity of particle types, not the statistical distribution of the types, which is covered in Fig. 5.

**Figure. 5** Statistical analyses of core-shell distributions of the Pt-Co nanoparticles from Fig 4. The particle populations, both before and after cycling in the fuel cell, follow log-normal distributions (Supplementary Fig S1). The colorbars in panel (**a**) relate the color of the markers

to the surface-area-normalized probability of a spectroscopically-measured particle's occurrence in an unbiased measurement. The Pt-rich, Co-free shell for the starting distribution of particles is almost independent of particle radius (measured by the effective Pt radius as shown in Supplemental Fig S10). After cycling, there is a wide spread in effective shell thickness, which increases as the particle size increases. The size of the marker depicts the number of distinct Co cores found in the particle, with larger markers signifying more cores. The dashed white line marks where the Pt-shell would extend throughout the particle, ie. a total absence of Co. The fraction of measured particles that contain multiple Pt-Co cores (**b**) increases with particle size, accounting for roughly half the particles with radii larger than 6nm. However, there are many more small particles than large, so surface area is dominated by the smaller particles (**c**). For understanding the impact of shell thickness on electrochemical activity, the fraction of surface area as a function of Pt-shell thickness is plotted in (**d**). After 30,000 voltage cycles (30k), the Pt-shell thickness has grown and is thicker for the multi-core particles than the single core. The data presented in (**c**) and (**d**) were normalized to the population distribution (see methods and Supplementary Fig S2) such that we were not biased by sampling preferences.

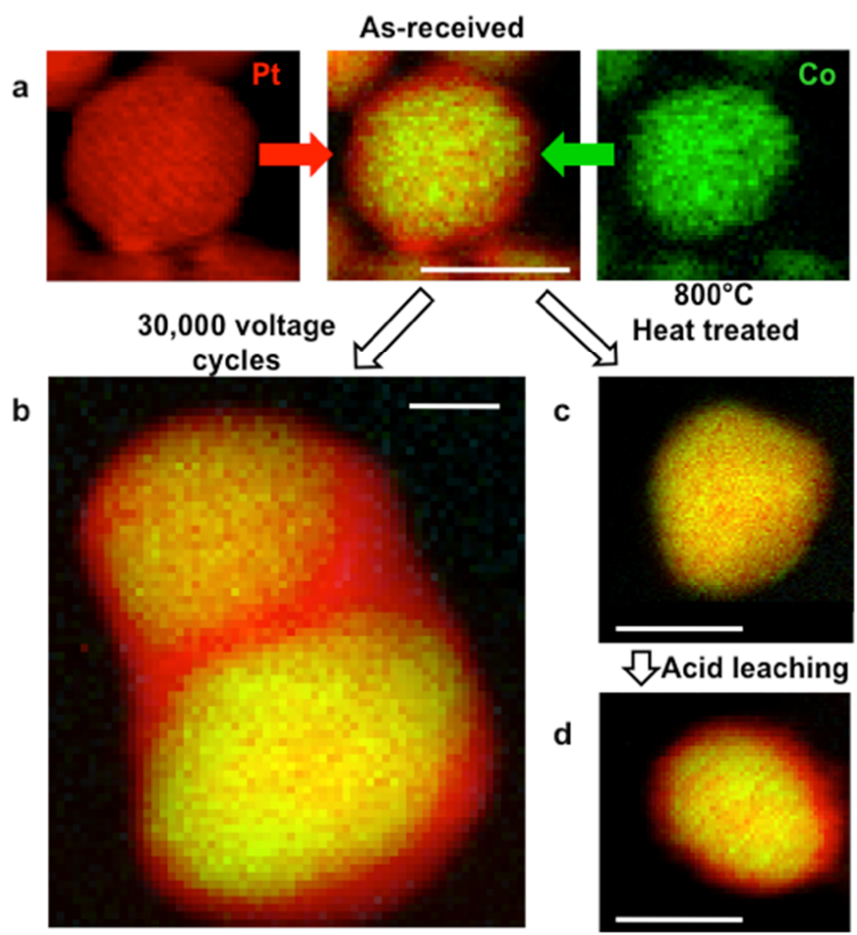

**Figure 1.** Typical EELS spectroscopic images of Pt-Co nanoparticles for the chemical and electrochemical processes studied here. The relative Pt concentration is shown in red and the Co concentration in green, so yellow indicates a Pt-Co alloy. The stripes in the Pt image of (**a**) are {111} lattice planes viewed at a slight mistilt. (**a**) as-received, showing 0.6 nm Co-free shell, (**b**) after 30k voltage-cycles, showing a coalescence of 2 smaller Pt-Co nanoparticles surrounded by a Co-free shell ~2-3 nm thick, (**c**) after heat-treatment of the as-received material, leading to a uniform Co distribution and no Co-free shell and (**d**) after acid-leaching the heat-treated sample, where a Co-free shell similar to that of the starting material (**a**) has returned. The scale bar for each image is 5 nm.

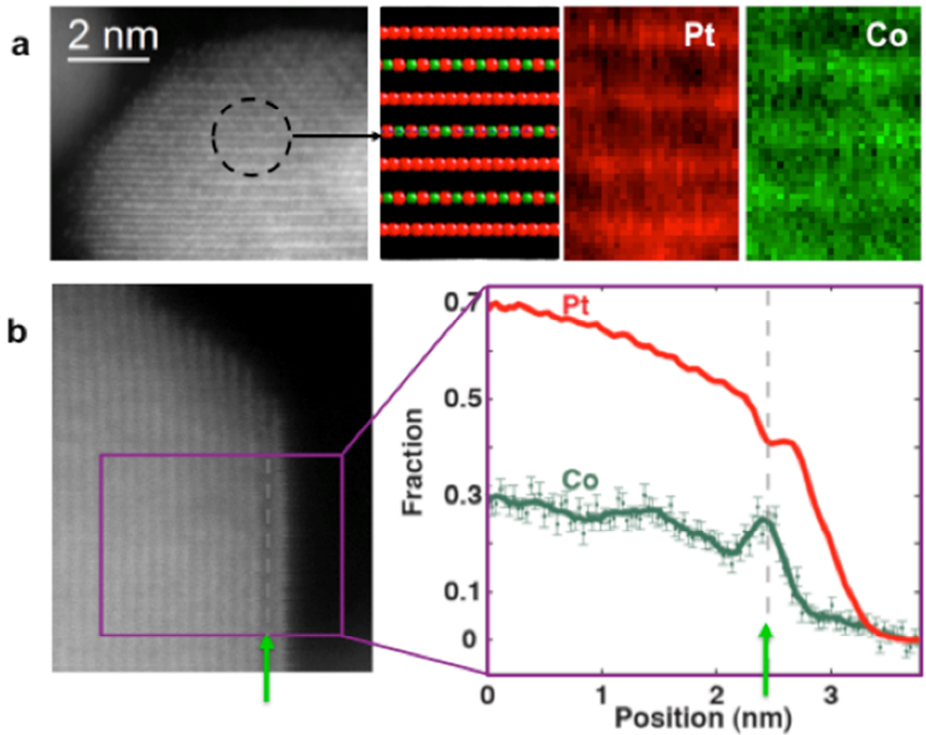

**Figure 2.** Atomic-resolution spectroscopic images of Pt-Co nanoparticles from the heat-treated sample. **a**, a Ll$_2$ ordered Pt-Co nanoparticle with (001) planes oriented parallel to the electron beam. The Pt and Co maps show the alternating planes of the Pt and Pt/Co as shown in the model structure (Pt - red, Co - green). **b**, a {111}-facet terminated Pt-Co nanoparticle, and the Pt/Co concentration profile across the facet showing a strong de-alloying effect at the last two planes, with the Co segregating to one layer below the surface, and Pt segregating to outermost plane. The Pt and Co signals are scaled to their nominal bulk compositions at x=0.

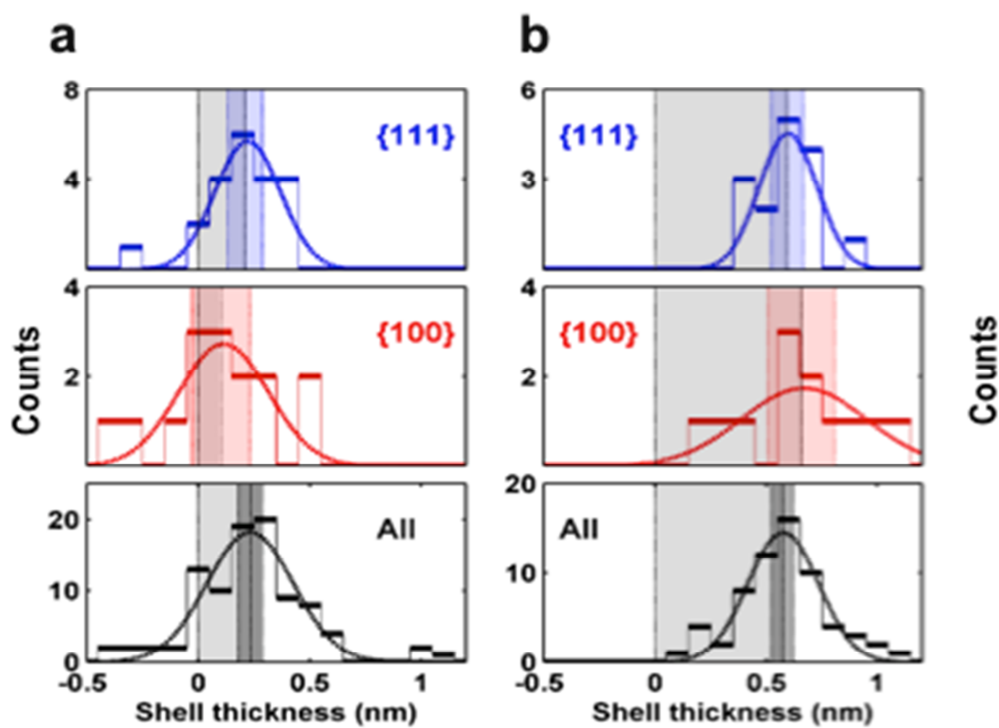

**Figure 3.** Histograms of the Pt shell thickness, measured from different facets of individual nanoparticles in (**a**) the heat-treated sample and (**b**) the acid-leached sample. The left and right boundary of the colored shaded area marks the two standard error of the mean (95.4% confidence interval) and the right boundary of the gray shaded area marks the average shell thickness. The overlap of the red and gray areas for the {100} facets of the heat-treated sample shows that there is no statistically-significant shell thickness there, whereas there is a 1.6 ± 0.5 Å (~1 monolayer) Pt shell on {111} heat-treated facets. In contrast, after acid treatment (**b**) the shell thickness has grown to 5.7 ± 0.3 Å (~ 2.5 monolayers) and is independent of facet orientation. For the heat-treated sample, of the 94 facets measured, 21 could be indexed as {111}, and 15 as {100} while for the acid-leached material, of the 63 facets, 15 were identified as {111} and 12 as {100}.

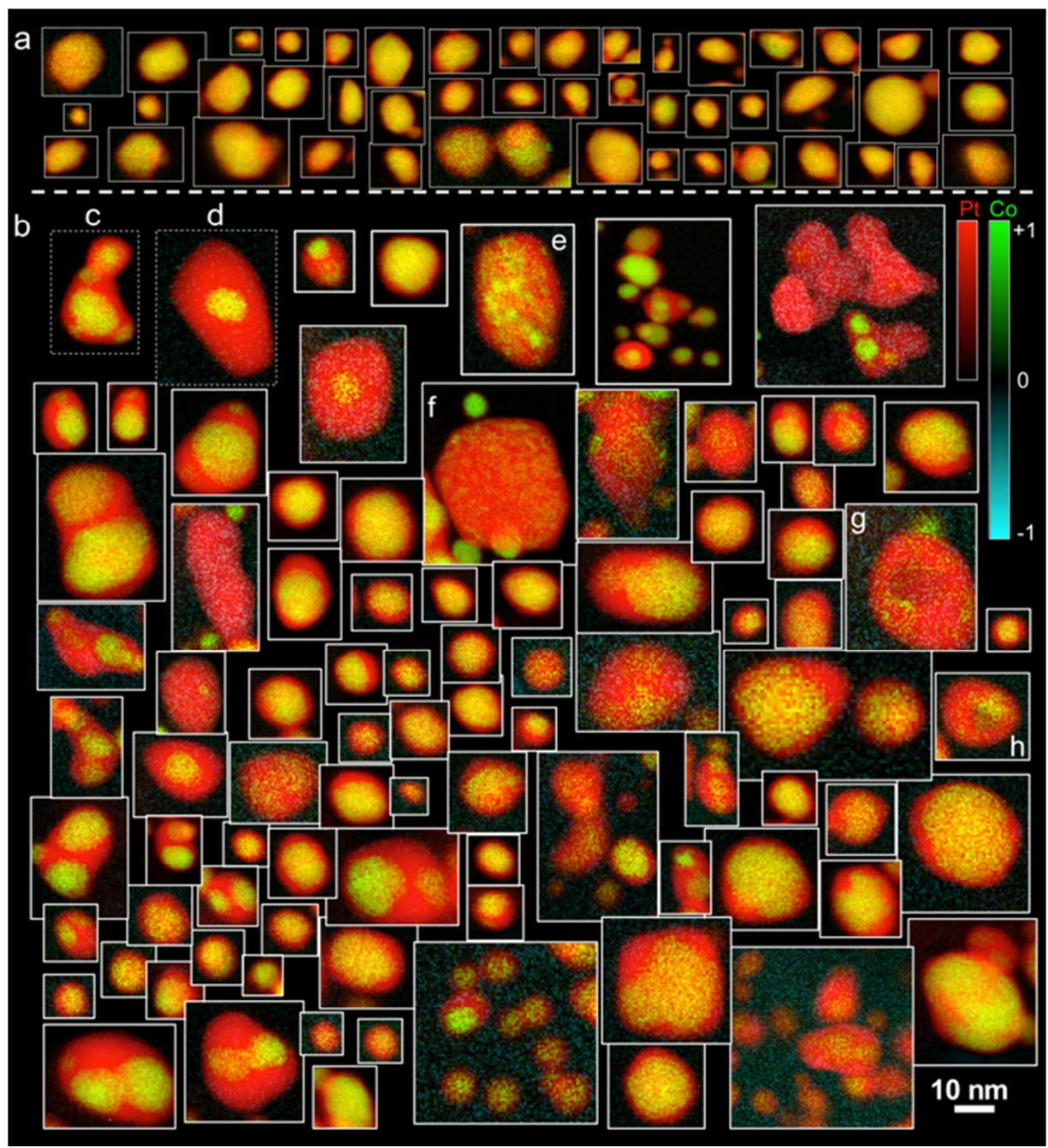

**Figure 4.** Collage of EELS spectroscopic images from the (**a**) as-received and (**b**) the voltage cycled sample. As shown in the color bars, the relative Pt concentration, normalized in each separate image, is plotted in red and the Co concentration ranging from green to turquoise. Particles (**c-h**) from the cycled sample are discussed in more detail in the text. The selection of images shown is meant to display the diversity of particle types, not the statistical distribution of the types, which is covered in Fig. 5.

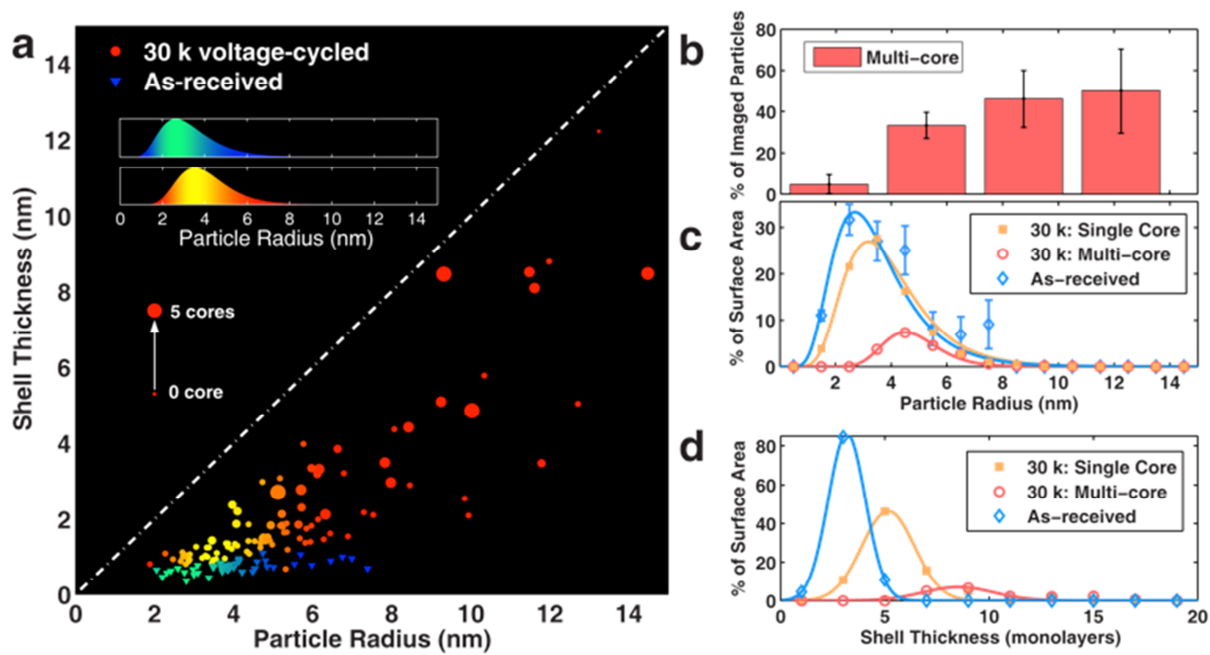

**Figure. 5** Statistical analyses of core-shell distributions of the Pt-Co nanoparticles from Fig 4. The particle populations, both before and after cycling in the fuel cell, follow log-normal distributions (Supplementary Fig S1). The colorbars in panel (**a**) relate the color of the markers to the surface-area-normalized probability of a spectroscopically-measured particle's occurrence in an unbiased measurement. The Pt-rich, Co-free shell for the starting distribution of particles is almost independent of particle radius (measured by the effective Pt radius as shown in Supplemental Fig S10). After cycling, there is a wide spread in effective shell thickness, which increases as the particle size increases. The size of the marker depicts the number of distinct Co cores found in the particle, with larger markers signifying more cores. The dashed white line marks where the Pt-shell would extend throughout the particle, ie. a total absence of Co. The fraction of measured particles that contain multiple Pt-Co cores (**b**) increases with particle size,

accounting for roughly half the particles with radii larger than 6nm. However, there are many more small particles than large, so surface area is dominated by the smaller particles (**c**). For understanding the impact of shell thickness on electrochemical activity, the fraction of surface area as a function of Pt-shell thickness is plotted in (**d**). After 30,000 voltage cycles (30k), the Pt-shell thickness has grown and is thicker for the multi-core particles than the single core. The data presented in (**c**) and (**d**) were normalized to the population distribution (see methods and Supplementary Fig S2) such that we were not biased by sampling preferences.

**Supplementary information for**

# Atomic-resolution spectroscopic imaging of ensembles of nanocatalyst particles across the life of a fuel cell

Huolin L. Xin[1,*], Julia A. Mundy[2,*], Zhongyi Liu[3], Randi Cabezas[4], Robert Hovden[2], Lena Fitting Kourkoutis[2], Junliang Zhang[3], Nalini P. Subramanian[3], Rohit Makharia[3], Frederick T. Wagner[3], and David A. Muller[2,5]

1. Department of Physics, Cornell University, Ithaca, NY 14853, USA
2. School of Applied and Engineering Physics, Cornell University, Ithaca, NY 14853, USA
3. Electrochemical Energy Research Laboratory, General Motors, Honeoye Falls, NY 14472, USA
4. Electrical and Computer Engineering, Florida International University, Miami, FL 33174, USA
5. Kavli Institute at Cornell for Nanoscale Science, Ithaca, NY14853, USA

* These authors contributed equally to this work.

## Contents

1. Supplementary methods
2. Legends for movies
3. Supplementary table S1- S2 and supplementary figure S1-S16
3. Supplementary references

**Supplementary methods**

**Material.** Pt$_3$Co/C catalyst supported on Vulcan® XC-72 carbon black with a Pt loading of approximately 30 wt.% (30%Pt$_3$Co/C) was received from Tanaka Kikinzoku Kogyo K. K. (TKK), Japan. The as-received 30%Pt$_3$Co/C catalyst underwent further heat- and acid-treatments as described below.

1. Heat-treatment: 0.5 g of the as-received catalyst was loaded in a quartz boat and annealed in a tube (inner diameter ~ 4.5 cm) furnace at 800 °C for 6 h. The annealing was done in a reducing atmosphere with mixed gas flow rates of 0.1slpm (standard liter per minute) of H$_2$ and 1.9 slpm of N$_2$. After being cooled down to room temperature in the same reducing atmosphere, the catalyst was purged with pure N$_2$ for 30 min; and then air was gradually introduced into the furnace tube before the catalyst was taken out.

2. Heat plus acid-treatment: 0.5 g of the heat-treated catalyst was soaked in 0.5 L of 1M HNO$_3$ in a 3-necked reflux flask, and treated at 80 °C in a heating mantle for 24 h. This acid- treated catalyst was then filtered and washed with plenty of Millipore® ultra pure de-ionized water. The catalyst was then dried in a vacuum oven at 60 °C for 4 h.

**Electron Microscopy.**

**Sample preparation:** The as-received, heat-treated, and heat-treated plus acid-leached samples were dispersed in isopropanol and dropped onto a holey carbon grid. The 30k-cycled sample was cut from the membrane-electrode-assembly (MEA) by ultra-microtoming to a sample thickness of ~ 30-50 nm.

**Atomic-resolution spectroscopic imaging**: The Nion UltraSTEM was operated at 100 kV in spectroscopic imaging mode with a 33 mrad convergence semi-angle, a typical current of 100-200 pA, spatial resolution of ~1-1.4 Å, and energy resolution of 0.35 – 1.0 eV. Using an additional corrector for improved coupling to the Gatan Enfina spectrometer, electrons scattered up to 77 mrad can be efficiently collected without sacrificing the energy resolution. We choose to operate the microscope at a relatively low beam voltage (100 keV) to avoid surface sputtering and mass loss of Co that can occur above ~120 keV[1]. The Co concentration in individual particles was mapped using a power-law background subtraction and integration window at the Co L$_{2,3}$ edges[2] (Supp. Fig. S6b). Both the Pt N$_3$-edge and the Co L$_{2,3}$-edge were recorded simultaneously, but longer than usual acquisition times (0.05-0.2 sec/spectrum) were needed to reliably discriminate the Pt N$_3$ edge from the large underlying background (Fig. S6, S7 and S14). An entire 64×64 pixel map could thus be collected in ~10 minutes and larger maps (e.g., 150×150 pixels) in ~40 minutes. As the Pt N$_3$ edge at 518 eV is broad and has a delayed onset, a similar method yielded a noisy Pt signal (Supp. Fig. S8c). A multivariate curve resolution (MCR) method[3-9] was used here to decouple Pt signal from background over a wide energy range (450-650 eV) (Supp. Fig. S7). Using this method, a 10-fold increase in signal-to-noise ratio was achieved compared to the results from power law background subtraction and it yields the same result as these of the ADF and the background subtraction methods (Supp. Fig. S8). While the Pt N$_3$ edge with its onset at 518 eV can already be distinguished from the nearby O-K edge at 532 eV (Supp. Fig S14b), least squares MCR fits which also exploit the very different shapes and positions of the two edges allow the energy range of comparison to be extended. This improves the signal-to-noise ratio and still allows maps of both Pt and O when both are present

(Supp. Fig S14). To further test the reliability of the MCR method, we compared the Pt-$N_3$-MCR map with the Pt map extracted from the Pt $M_{4,5}$ edge at 2200 eV where there is no possibility of overlap with the O-K edge (Supp. Figure S13). Supplementary Figure S13 shows that the Pt-$N_3$-MCR map is almost identical to that of the Pt $M_{4,5}$.

**Shell thickness calculation**: To determine the Pt shell thickness along a specified facet, the particle was indexed using the diffractogram. The Pt and Co line profiles drawn through the terminating facet were fit with the sum of two error functions and the difference in the full width at half maximum compared (Supplementary Figure S9). For the effective shell thickness for the entire particle, the Co and Pt areas were found by counting all the pixels that fell above a threshold (50% of the maximum signal intensity for each element). The particle and the core were then modeled as a sphere to compute the radius from the area; the difference between the Pt radius and Co radius was used as the shell thickness (Supplementary Figure S10).

**Statistical sampling**: For the as-received material, 29 out of 75 EELS maps were excluded from further processing because the Co map intensities (but not shapes) were distorted by multiple scattering on very thick portions of the support and could not be reliably quantified although general qualitative features could still be discerned. These were not included in the either qualitative or quantitative discussion in the paper, but are given Supplementary Figure S15, where the reader can see they were similar in structure to particles sampled on the thinner regions of the support.

**Electron tomography**: The 3-D tomogram as shown in Supplementary Figure S5 and Supplementary Movie S1 was reconstructed from 97 ADF-STEM images taken from -72 to +72 degrees with 1.5-degree intervals acquired in a FEI Tecnai F20 Field Emission Gun with Fischione annular dark field detector operated at 200 kV (probe current ~10pA; probe-forming angle 10 mrad; nominal camera length of 300 mm, recording scattered electrons from 22 mrad to 110 mrad).

**Rotating disk electrode test.** A RDE-3 Potentiostat (Pine Instrument Company, USA) was used to conduct rotating disk electrode (RDE) studies. A glassy carbon disk (~ 0.5 mm in diameter), a Pt-wire, and a calomel electrode were used as the working, counter, and reference electrodes respectively with the reference electrode in a separate compartment connected via a salt bridge. All potentials are reported against the reversible hydrogen electrode (RHE) in the working electrolyte, with the correction factor checked periodically by measuring the current onset of $H_2$ oxidation in a hydrogen-sparged electrolyte. To make the thin film RDE (working electrode), the electrode ink was prepared by the addition of 15 ml of $H_2O$ : IPA (isopropyl alcohol) in the volume ratio 4 : 1 to the catalyst (1 mg catalyst / 1 ml solvent), followed by the addition of 60 μl of 5 wt.% Nafion® SE5112 (DuPont, USA). This ink was ultrasonically dispersed and loaded on the glassy carbon RDE (to result in 15-25 $\mu g_{-Pt}/cm^2_{-RDE}$) and dried at room temperature under $N_2$. All electrochemical measurements were done in 0.1 M $HClO_4$ prepared from GFS doubly-distilled concentrated $HClO_4$ (GFS Chemicals, USA). Before measurement of Pt-surface area and ORR activity, the electrode was conditioned by cycling in $N_2$-purged 0.1 M $HClO_4$ at 1 V/s between ~ 0.025 and 1.2 V. For measurement of Pt-surface area, cyclic voltammograms (CVs) were recorded in $N_2$-purged 0.1 M $HClO_4$ at 20 mV/s between ~ 0.025 and 1.2 V - and the hydrogen-adsorption-desorption (HAD) area integrated and converted to real Pt surface area assuming 210 $\mu C/cm^2_{Pt}$. For measurement of ORR activity, linear sweep voltammograms

(LSVs) were recorded in $O_2$-purged 0.1 M $HClO_4$ at 5 mV/s as the positive-going sweep between ~ 0.025 and 1.2 V. Activities for catalysts are stated as kinetic currents at 0.9 V.

**Fuel cell test.** 50 $cm^2$ catalyst-coated membranes (CCMs) were used for fuel cell tests. The as-received 30% $Pt_3Co/C$ was used as the cathode catalyst. The anode catalyst was 20 wt.% Pt/Vulcan® (TKK, Japan). Nafion® DE2020 (DuPont, USA) was used as the ionomer in both the cathode and the anode. Nafion® NRE211 was used as the membrane. The gas diffusion media used were proprietary. The Pt surface area (i.e., HAD) of the cathode electrode was measured from CVs at room temperature, sweep rate of 20 mV/s, 3 cycles from 0 mV to 600 mV *vs.* RHE. Catalyst activity was measured from HFR (high frequency resistance)-corrected currents at 0.9V from negative sweep 'pure $O_2$' polarization curves. The $O_2$ polarization curves were measured at 100% relative humidity (RH), 80°C, 150 $kPa_{abs}$ and 2 / 10 stoichiometry of $H_2$ / $O_2$ on anode / cathode. Detailed test procedures can be found elsewhere[10]. In a durability test, the degradation of the cathode catalyst was accelerated by voltage cycling for 30,000 cycles between 0.61 V and 1.01 V *vs.* RHE at a sweep rate of 50 mV/s. The durability test was carried out at 80 °C, 101 $kPa_{abs}$, 100% $RH_{inlet}$, 200 sccm (standard cubic centimeters per minute) / 50 sccm of $H_2$ / $N_2$ on the anode / cathode.

**Legends for supplementary movies**

**Movie S1.** A 360-degree view of the three-dimensional (3-D) tomographic reconstruction of the as-received Pt-Co nanoparticles on the carbon support as shown in Supplementary Figure S5.

**Supplementary tables and figures**

|  | As -received | Heat-treated | Heat plus acid-treated | 10k voltage cycled | 30k voltage cycled |
|---|---|---|---|---|---|
|  | $Pt_3Co/C$ | $Pt_3Co/C$ | $Pt_3Co/C$ | $Pt_3Co/C$ | $Pt_3Co/C$ |
| HAD ($m^2/g_{Pt}$) | 48.4 (32) | 37.0 | 36.0 | (23) | (18) |
| Mass activity ($A/mg_{Pt}$) | 0.185 (0.115) | 0.155 | 0.130 | (0.065) | (0.041) |
| Specific activity ($\mu A/cm^2_{Pt}$) | 382 (350) | 422 | 358 | (285) | (240) |

**Table S1.** HAD, mass and specific activities of 30% $Pt_3Co/C$ measured by RDE and MEA (values in parentheses). Activities are lower than typical for Pt-alloy catalysts[2] because of the Vulcan carbon support, which is not optimum for alloy activity.

|  | As -received | 30k voltage cycled |
|---|---|---|
|  | Pt/C | Pt/C |
| HAD ($m^2/g_{Pt}$) | 56 (51) | (10.7) |
| Mass activity ($A/mg_{Pt}$) | 0.131 (0.084) | (0.023) |
| Specific activity ($\mu A/cm^2_{Pt}$) | 233 (165) | (212) |

**Table S2.** HAD, mass and specific activities of pure Pt nanoparticles on Vulcan carbon measured by RDE and MEA (values in parentheses)

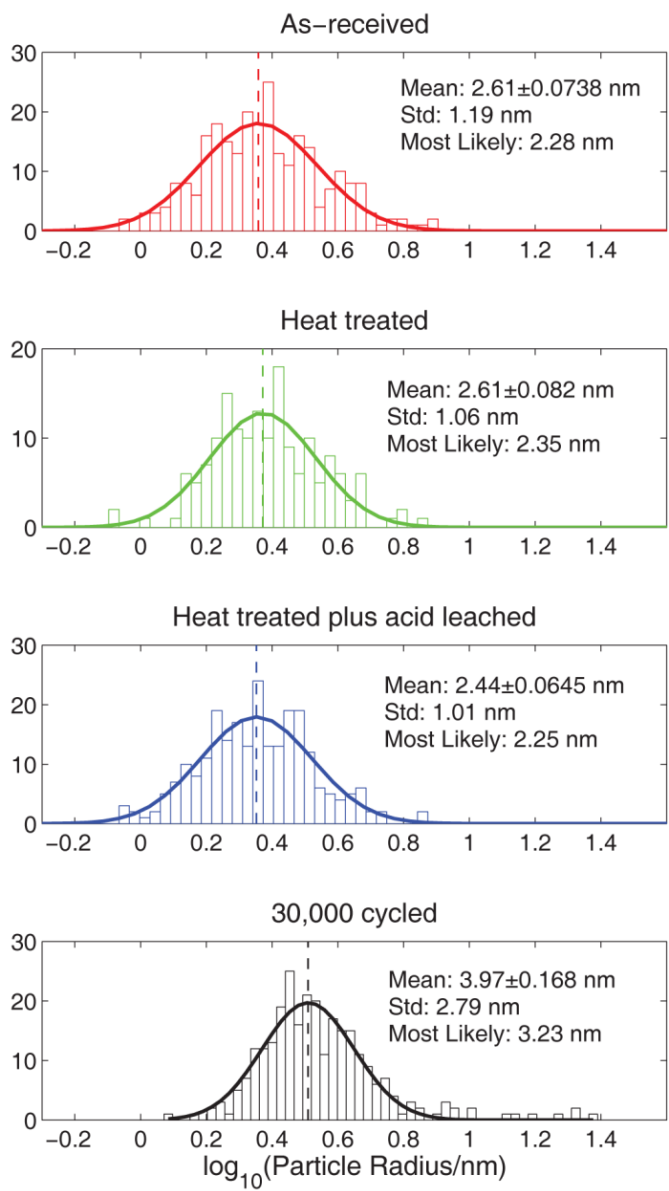

**Figure S1.** The particle size distributions of (a) as-received, (b) heated, (c) heated plus acid leached and (d) 30,000 cycled samples. The data is fit to a log-normal distribution. The mean, standard deviation (std) and the most likely particle sizes are given for each sample.

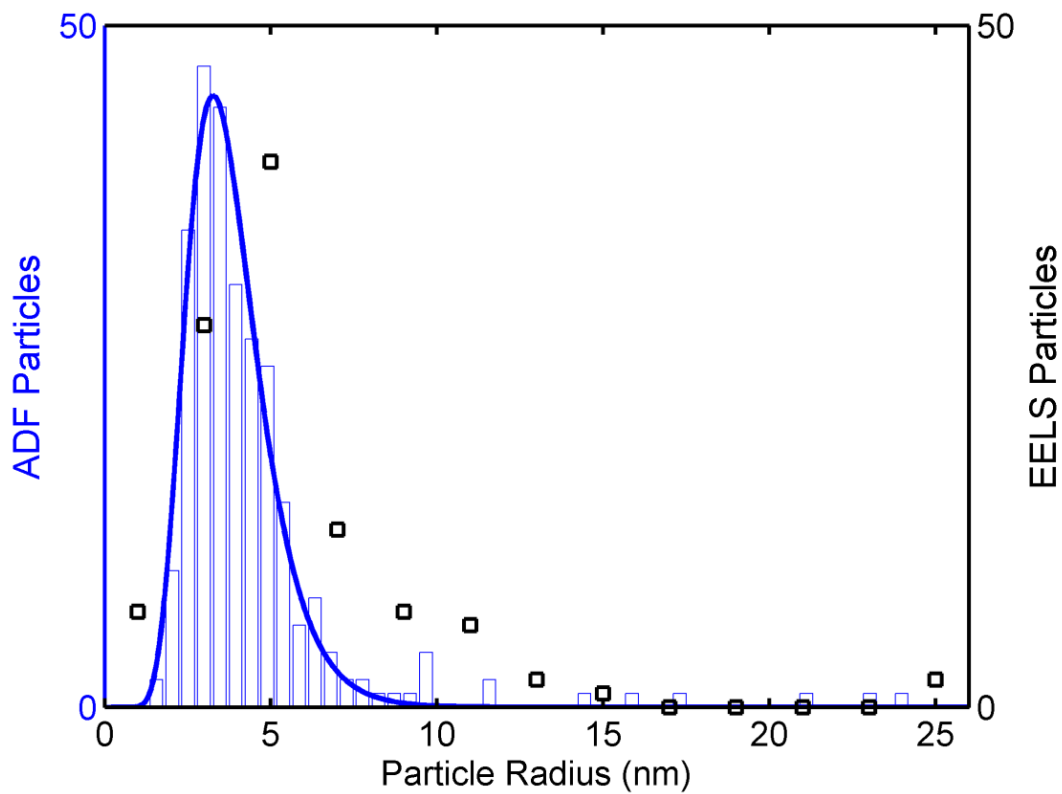

**Figure S2.** A comparison of the population (from the particle size distribution in Supplemental Fig S1) and EELS spectroscopic imaging selection for the 30,000 cycled sample. The ADF images are uniformly sampled so the ADF distribution is an accurate reflection of the nanoparticle size distribution. With a smaller sample size for EELS mapping, the larger particles were oversampled to develop better counting statistics on the coarsening mechanism. When computing statistics on the sample however, the sampled EELS data was weighted to the ADF population distribution to correct for our sampling bias.

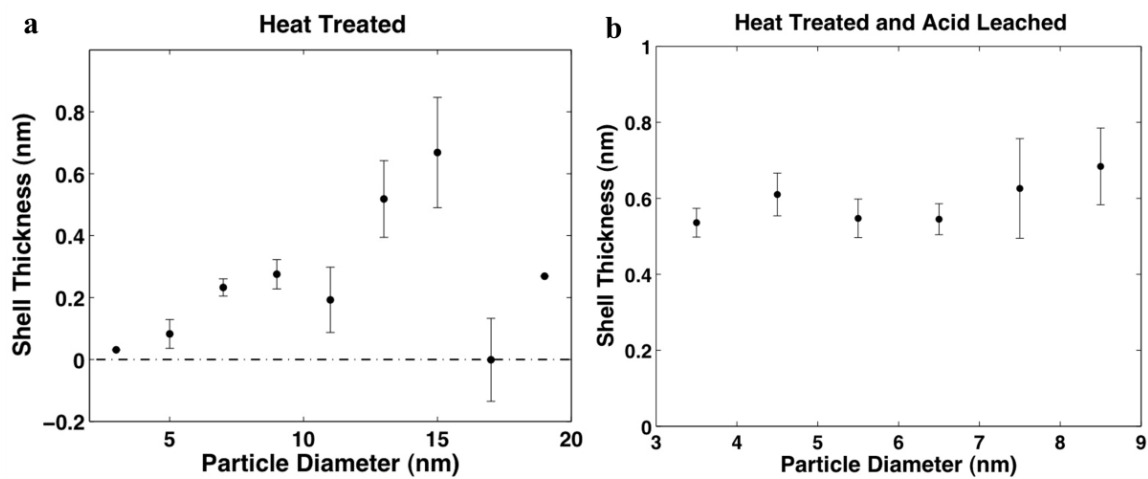

**Figure S3.** Plot of the Pt shell thickness vs. particle size of the Pt-Co nanoparticles (a) after heat treatment and (b) after heat plus acid-treatment. The vertical error bars represent the standard error of the mean. Note that a negative "Pt shell" thickness indicates the presence of a surface layer of Co. For both distributions, the shell thickness is independent of the particle size.

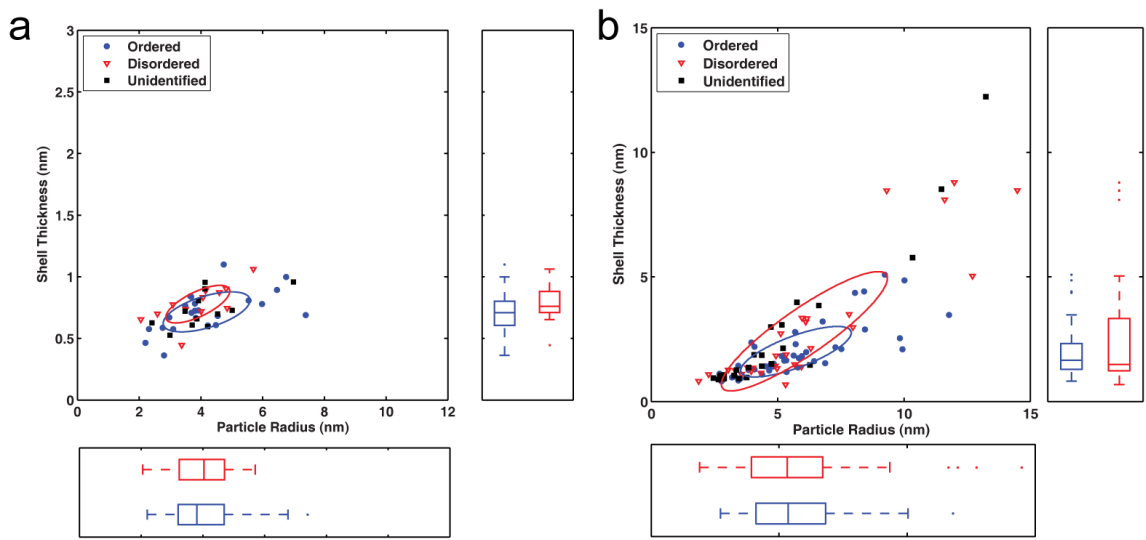

**Figure S4.** A comparison of the Pt shell thickness as a function of Pt radius for (a) the as-received sample and (b) the 30,000 cycled sample. The particles are grouped by the presence or absence of the Ll$_2$ ordering determined by the presence of a superlattice {110} or {100} peak in the diffractogram of the particle. Some particles were lying on the support such that the crystallographic axis too far off the axis of the electron beam for the diffractograms to be indexed; these particles are shown as "unidentified". The bi-variate normal ellipses (P=0.95) are also included. The box plots show that there is no statistically significant difference in either the particle size or the shell thickness for the ordered versus disordered particles for both the as-received and 30,000 cycled samples.

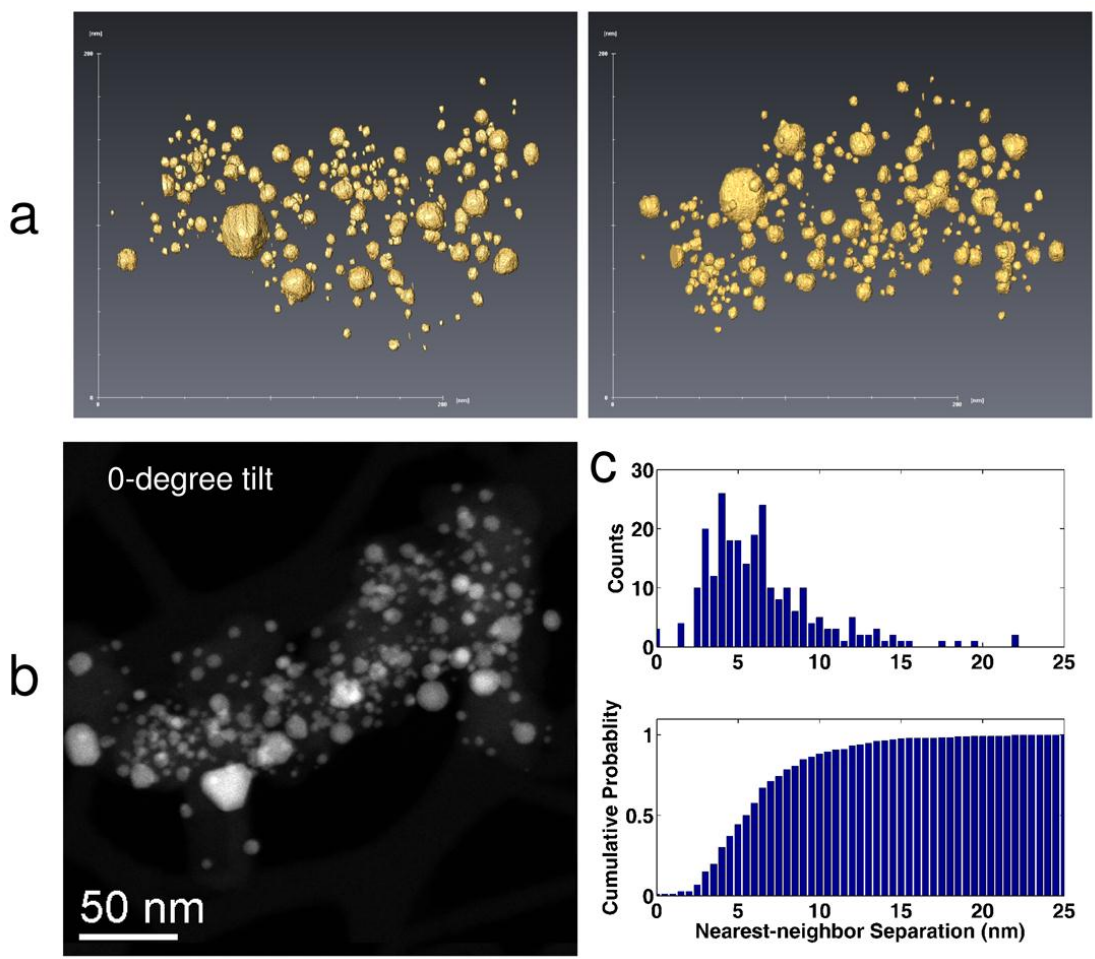

**Figure S5.** ADF-STEM tomographic reconstruction of the as-received Pt-Co nanoparticles on the carbon support was used to determine the inter-particle distances before voltage cycling. This serves as a baseline to differentiate between particles that coalesced during voltage cycling and those that had agglomerated before. (a) An isosurface visualization of the segmented particles from two different viewing directions demonstrating qualitatively that almost all particles are separated from their nearest neighbors prior to voltage cycling. Specifically, of the 249 particles in the reconstruction, only 3 were touching, an order of magnitude less than the number of particles which become part of a multi-core particle from the spectroscopic image analysis in Fig 5. A full movie is shown in Supplementary Movie S1. (b) The ADF-STEM image of the sample taken at 0-degree tilt. It appears in this projection image that many particles are touching, however from the tomographic reconstruction in (a) it is clear the apparently overlapping particles are at different z-heights and not touching. (c) A statistical quantification of the surface-surface separation between nearest particles. A histogram of each particle's nearest-neighbor separation is shown in the upper panel of (c) and the cumulative probability is shown in the lower panel of (c). From the analysis in Fig. 5, approximately 10% of the initial particles are found in a multi-core particle at the end of the lifetime. However from initial distribution in panel (c), even the closest 10% are separated by 2.5-3 nm from their neighbors, which is too far to produce in the multi-core particles observed in Fig 4b. Thus we conclude that the particles had to have moved on the support during cycling to produce the multi-core particles observed.

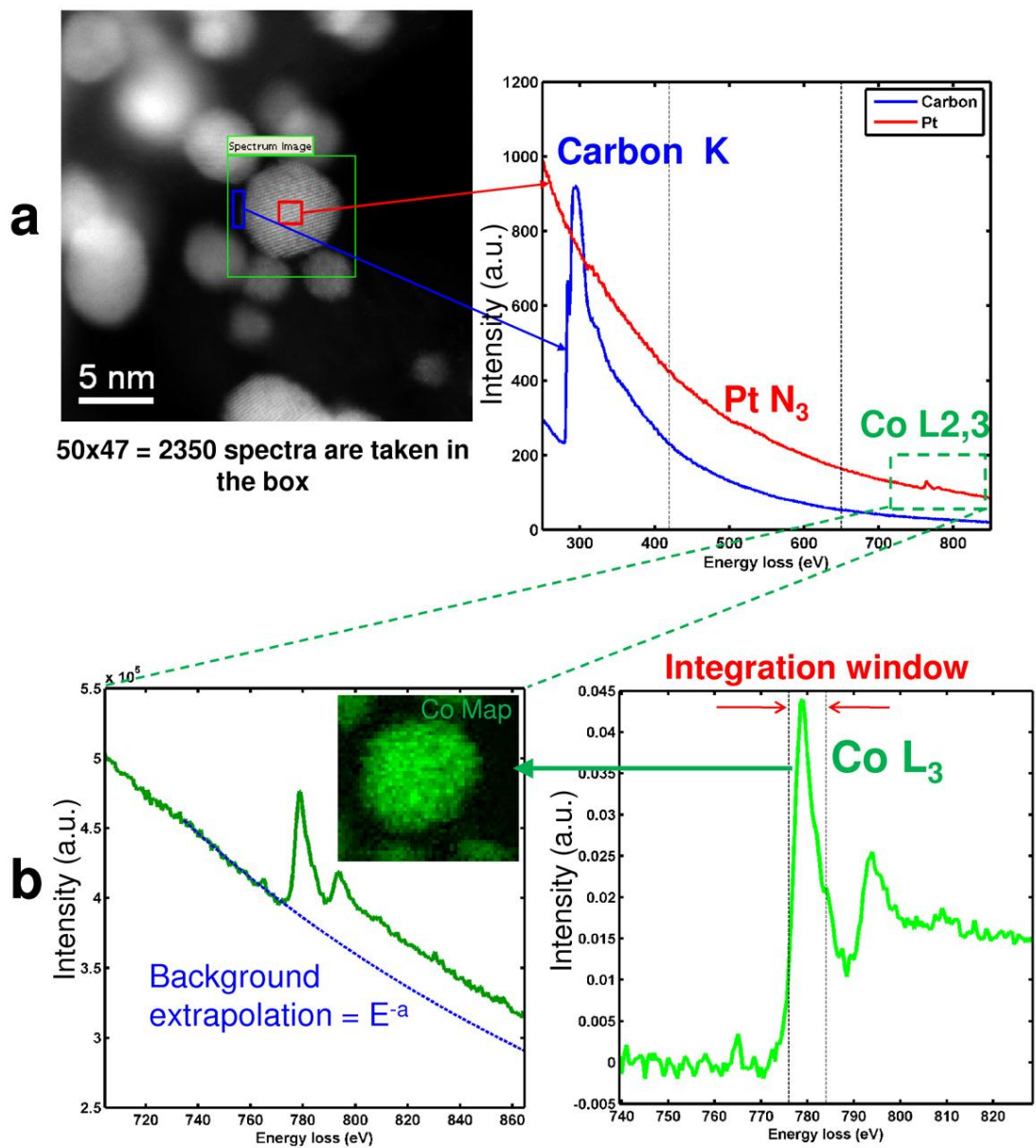

**Figure S6.** The extraction of the Co $L_{2,3}$-edge using a power law background subtraction method. (a) A comparison of the sample spectra from the center of the particle and the carbon support, clearing indicating the presence of the Co $L_{2,3}$ edge near 780 eV and the Pt $N_3$-edge at 518 eV as discussed in Supplementary Fig S7. (b) The fit of a power law background to a sample Co spectra and the resulting signal from which a narrow window (≈10 eV) is selected. The integration of the signal over this narrow window yields the Co map as shown in the inset.

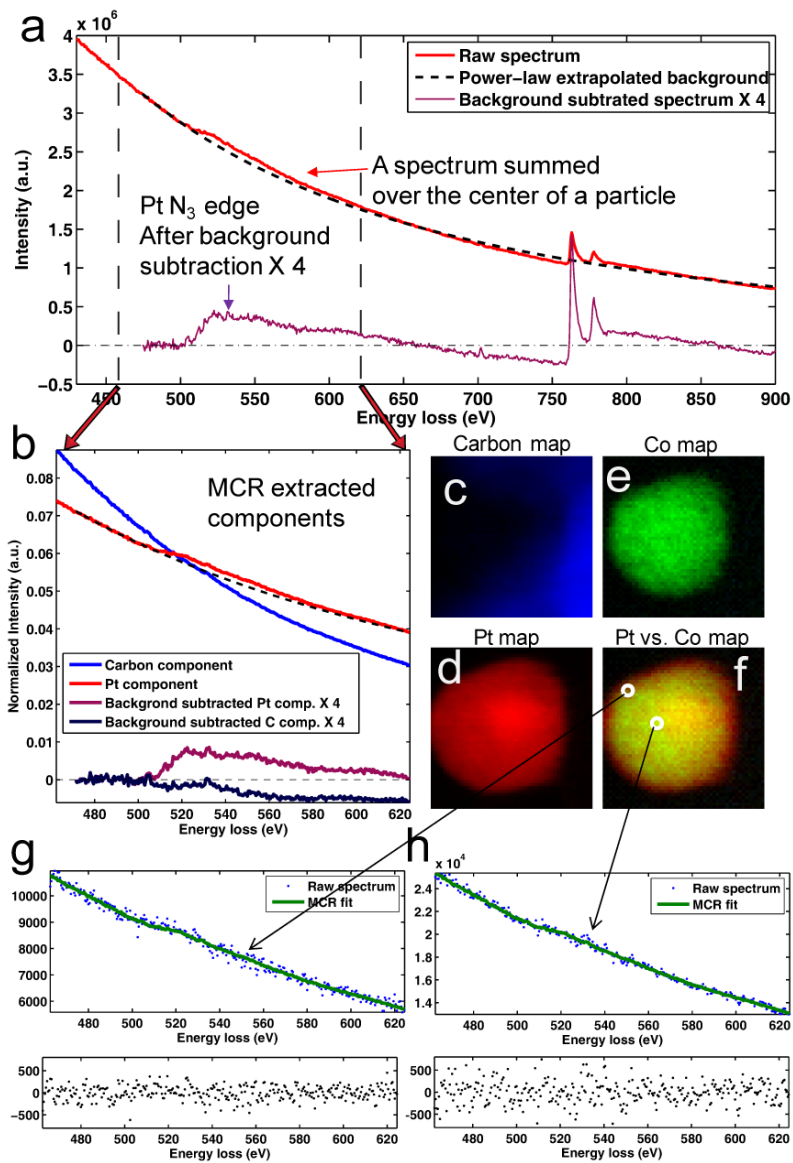

**Figure S7.** The process of applying multivariate curve resolution (MCR) [5, 11, 12] to the refinement of reference spectra for Pt and carbon. (a) A raw spectrum summed over the center of a particle. A power-law background fit (dashed line) only provides a good extrapolation for the Pt N$_3$-edge (~518 eV) over a very limited energy range, motivating our use of MCR to extend to usable energy window and enhance the signal. The background subtracted data (multiplied by four) is show at the bottom of (a). The minor Pt N$_3$ edge is broad and has a delayed onset. The peak-to-background ratio is intrinsically low and the power-law extrapolation crosses the spectrum at ~670 eV. Therefore using extrapolation-based method could render noisy and unreliable extrapolations (see the example in Supplementary Fig. S8). (b) The extracted Pt and carbon spectral components following 50 MCR iterations. (The power-law background subtracted data are again multiplied by 4 before plotting.) The background-subtracted Pt component reproduces the Pt N$_3$ edge as shown in (a). The background-subtracted carbon component shows a negative features indicates extended fine features are present in this energy range. The C and Pt concentration maps corresponding to these extracted spectra are shown in (c) and (d) respectively. (e) The cobalt concentration map. To demonstrate the quality of the fit,

two points were chosen from the composite map in (f), one in the center of the particle and the other in the shell. (g and h) plot the raw spectrum and that generated from the two-component fit for the central point and the point on the shell respectively. The residuals are shown below in black, demonstrating the goodness of the fit over the entire energy range used.

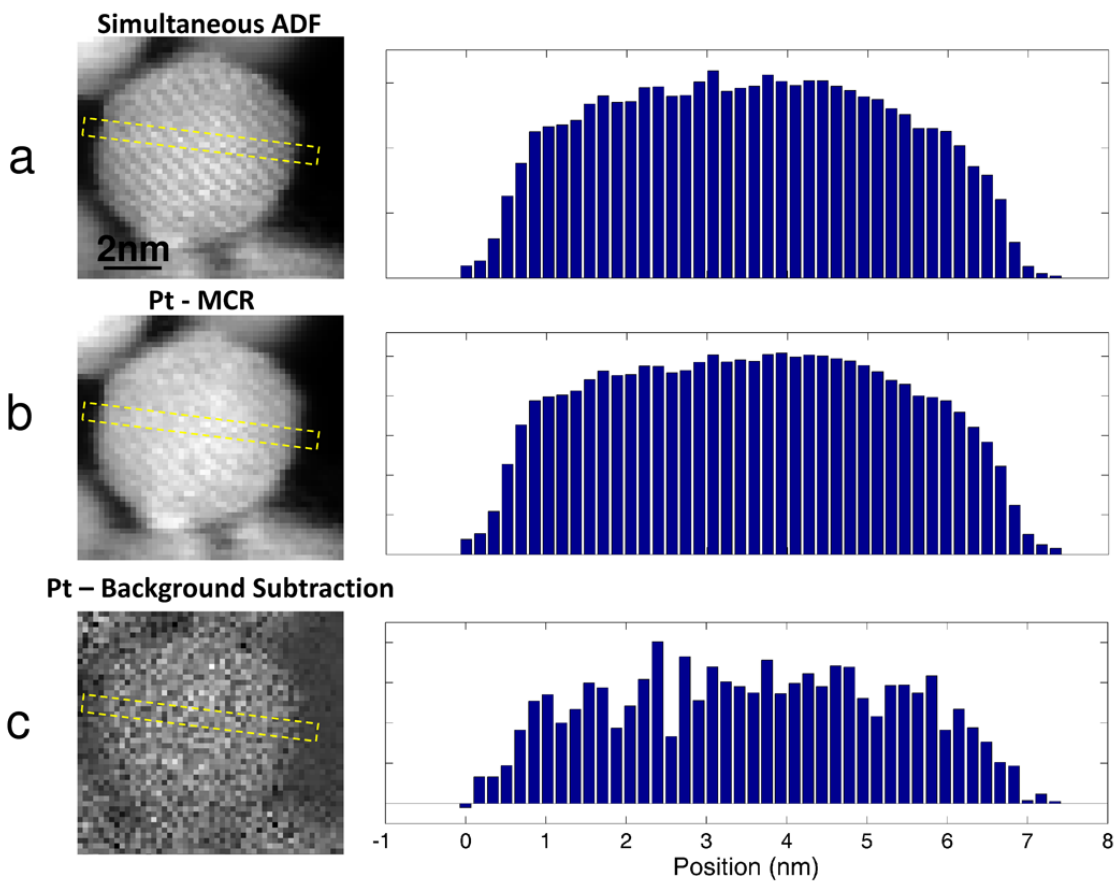

**Figure S8.** Comparison of methods that could be used to compute the particle size: (a) the simultaneously recorded annular dark-field (ADF) signal, (b) the MCR extracted Pt signal, and (c) the Pt signal obtained from a traditional power-law background subtraction and integration over the signal range. As the elastic ADF signal (a) is strongly dominated by the scattering from the heavy atoms, it tracks the Pt concentration$^2$. As shown in Fig S6, the MCR method exploits the unique scattering characteristics over a wide energy range to produce the Pt signal (b) with a high signal to noise ratio. The atomic lattice fringes present in the ADF signal are reproduced in the MCR Pt signal. Finally, the traditional background subtraction renders noisy maps as the peak-to-background ratio of Pt $N_3$ edge is poor (Supplementary Fig. S7a). The corresponding line profiles demonstrate that the particle size is consistently measured from all three of these signals on this nanoparticle. In this paper, the nanoparticles we studied are sitting on a semi-spherical carbon support. In some maps the particles are hanging off the edges. In this situation, the background ADF signal is uneven (from vacuum to support). Therefore, a quantitative measurement (Fig. S8) of the particle diameter using the ADF map could render unreliable results. Therefore we chose the MCR Pt signal for our analysis for its improved signal-to-noise ratio and reliability.

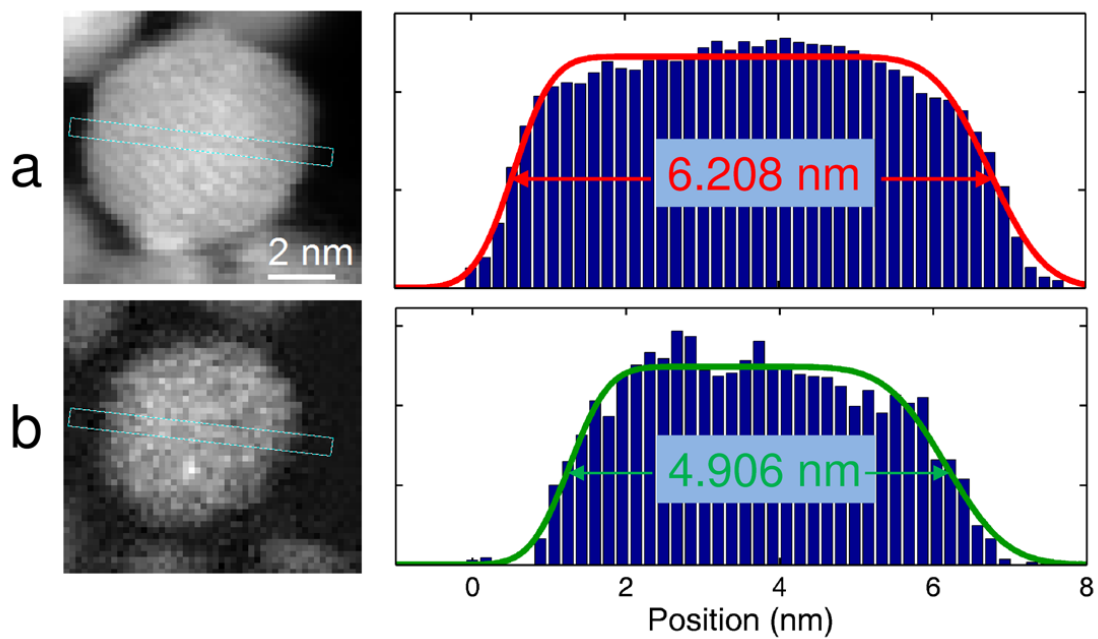

**Figure S9.** The fit of the (a) Co and (b) Pt signals with two error functions, $b \frac{E[a_1(x-x_1)] + E[a_2(x-x_2)]}{2}$ with fitting parameters $b$, $a_1$, $a_2$, $x_1$ and $x_2$, across a specified termination facet. As shown for this as-received particle, the width of the Pt signal is considerably larger than Co, indicating the presence of a Pt rich shell.

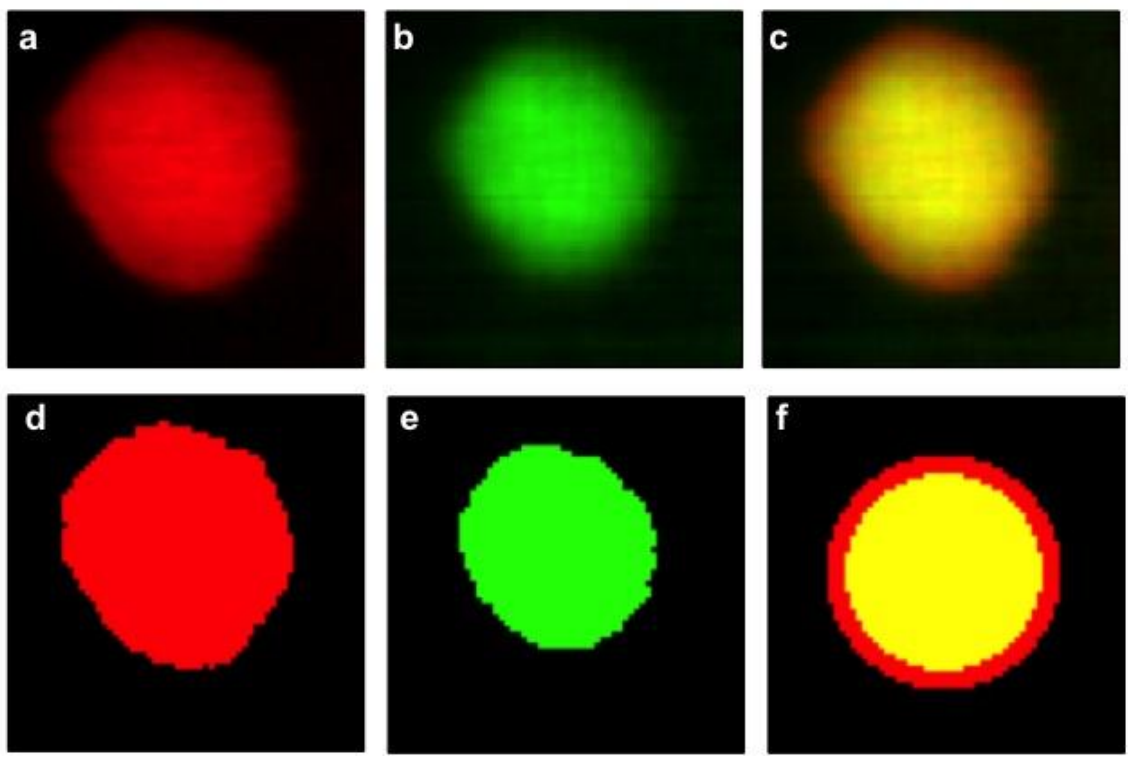

**Figure S10.** A sample analysis showing how the effective shell thickness and Pt radius were computed. The Pt, Co and composite signals for a typical as-received particle is shown in (a-c) respectively. The pixels lying above the threshold--50% of the maximum elemental signal intensity--for Pt and Co as used to compute the Pt area, $area_{Pt}$, and Co area, $area_{Co}$, are shown in (d) and (e) respectively. (f) A spherical model of the particle where the radius of the Pt signal, $r_{Pt}$, plotted in red is set by $r_{pt} = \sqrt{(area_{Pt}/\pi)}$ and the radius of the Pt-Co core l, $r_{Co}$, plotted in yellow is set by $r_{co} = \sqrt{(area_{Co}/\pi)}$.

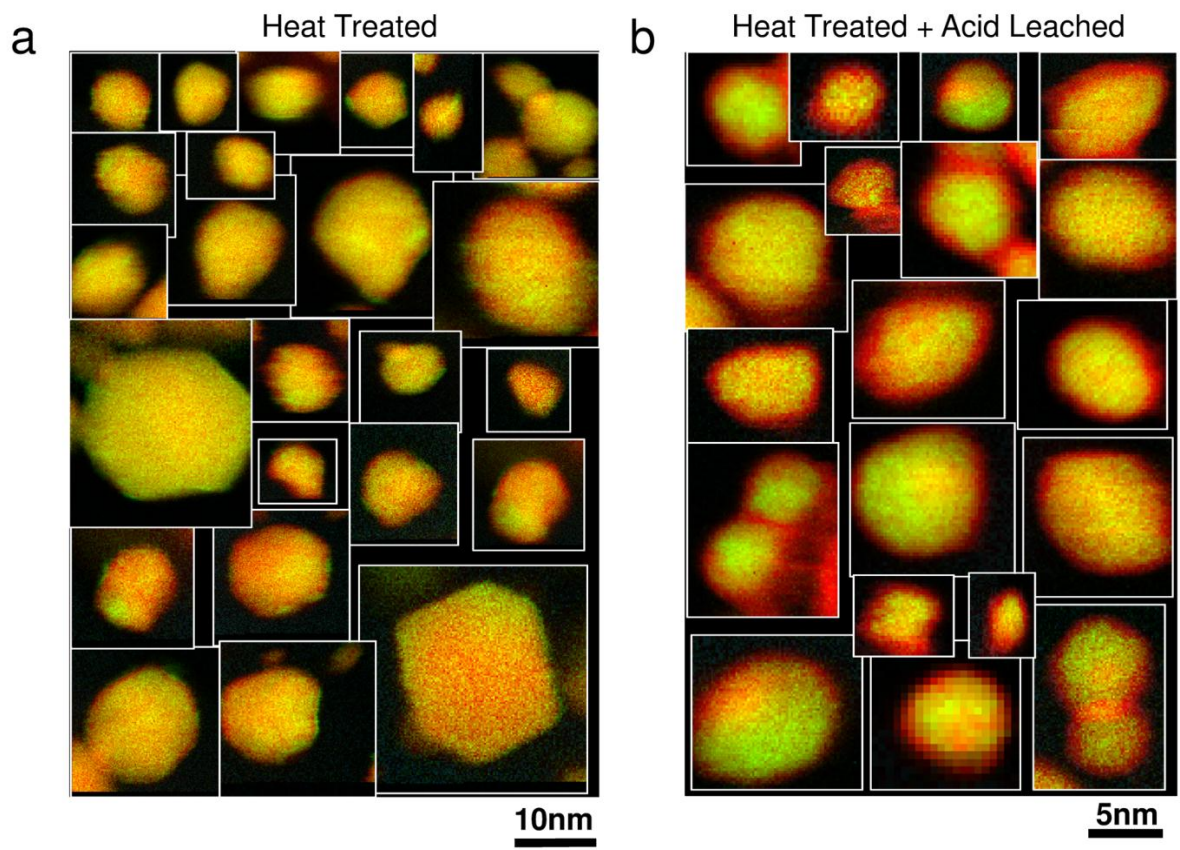

Figure S11. Spectroscopic maps from the heat-treated (a) and heat-treated and acid leached (b) particles.

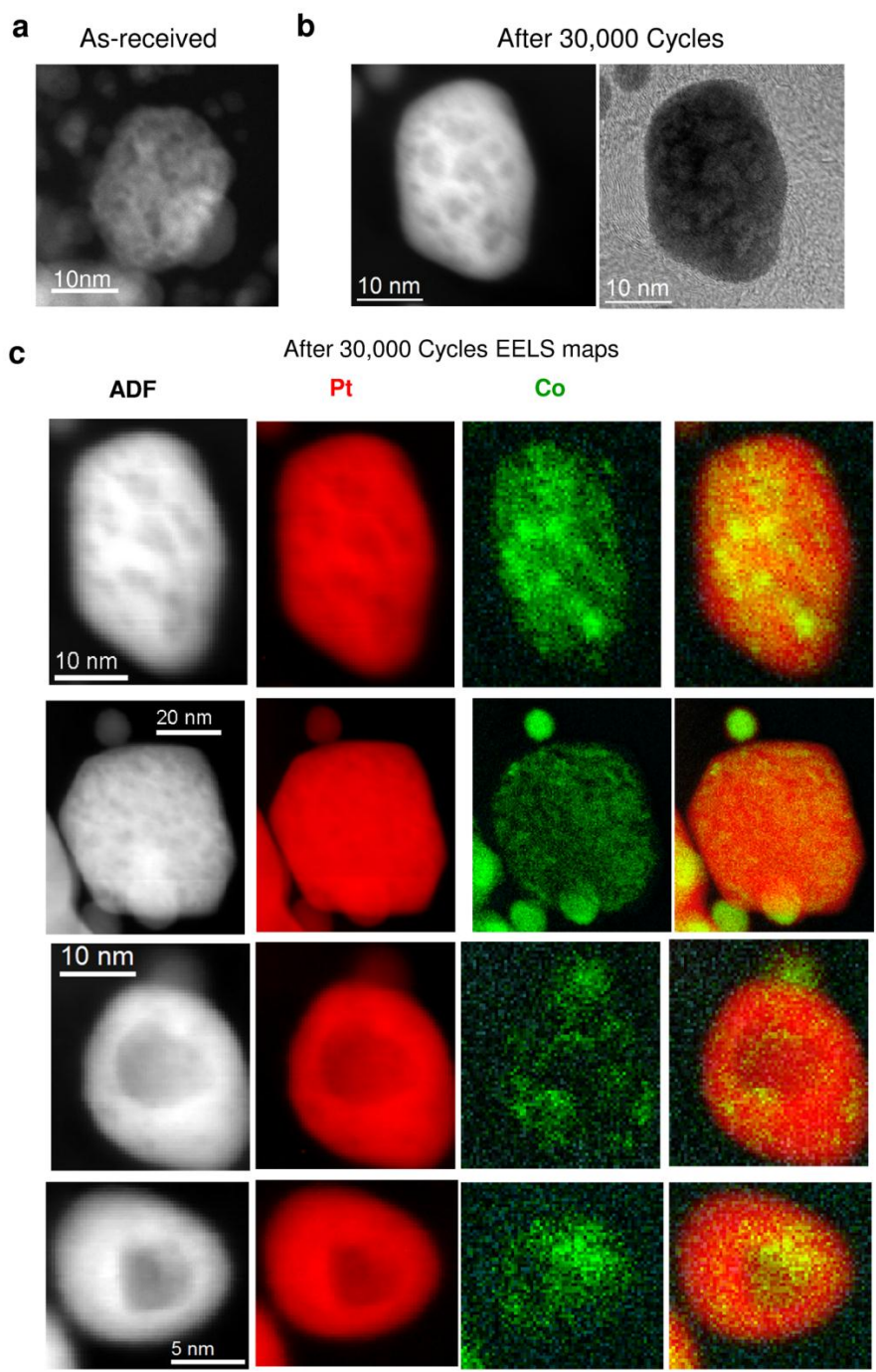

**Figure S12.** Particles with the "spongy" structure were observed both prior to and following the voltage cycling. An ADF image from the as-received material (a) and a BF and an ADF image following 30,000 voltage cycles (b) are shown. (c) The ADF image, Pt concentration, Co concentration and combined red-green image for the particles considered in Figure 4 e-h. From the ADF images, it is not clear whether the particles have voids or Co rich regions; this ambiguity is resolved with the Co maps.

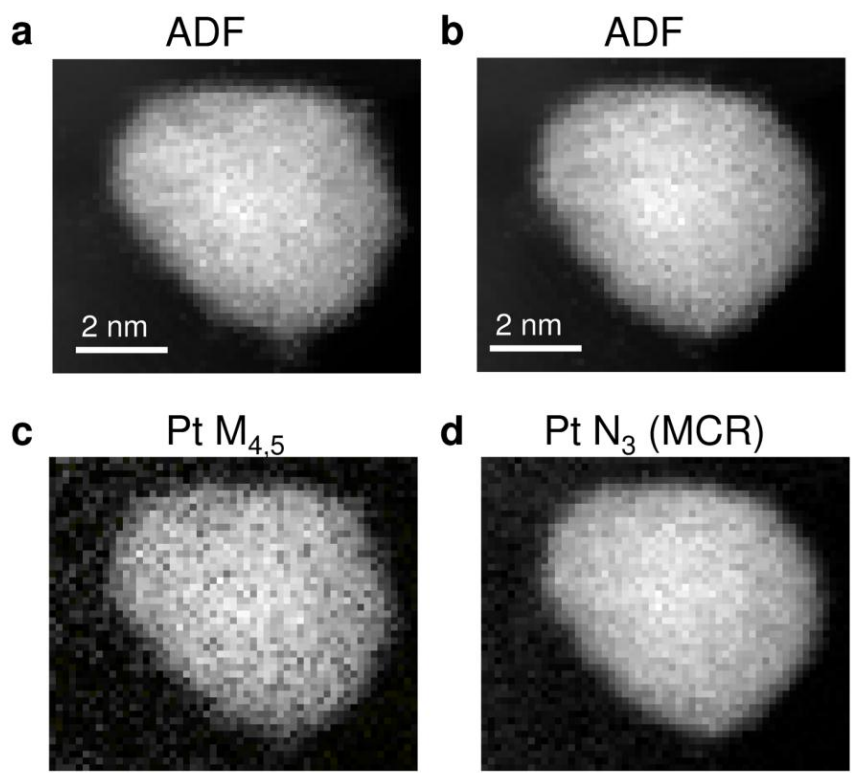

**Figure S13**. To demonstrate the robustness of the MCR extraction of Pt from the Pt $N_3$ EELS edge, two spectroscopic maps were recorded in succession from a single particle. The (c) background subtracted Pt $M_{4,5}$ edge at ~2200 eV and (d) MCR extracted signal from the Pt $N_3$ edge at ~518 eV are shown along with the simultaneously recorded ADF images, (a) and (b) respectively. While the two maps yield the same profile, the Pt $N_3$ edge was used in the analysis as it permitted the Co $L_{2,3}$ edge at 780 eV to be recorded simultaneously.

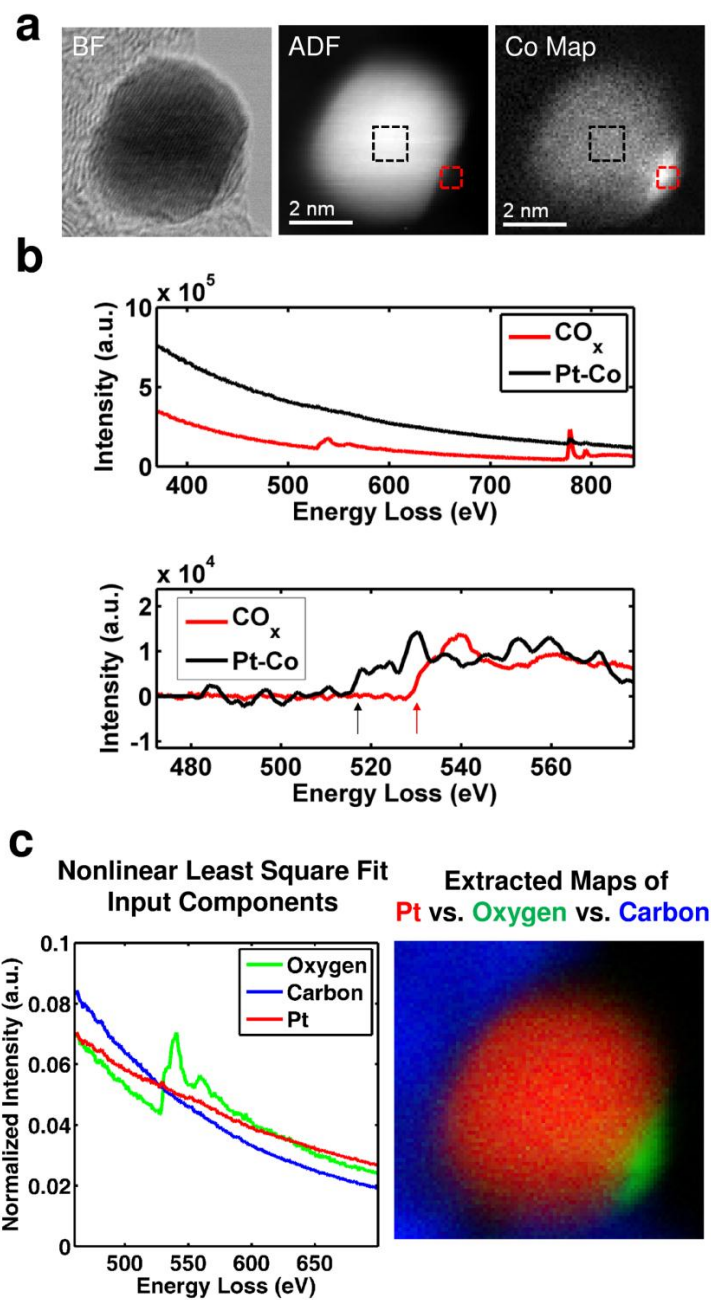

**Figure S14**. The Pt $N_3$ edge with an onset at 518 eV can be distinguished from the O K-edge at 532 eV when both are present. (a) The BF image, ADF image and Co map for a particle with a region of Co oxide near the surface. (b)The mean spectra from the center of the particle, where Pt-Co is present, compared to the Co-O rich region at the edge of the particle; both the raw spectra and background-subtracted spectra are displayed to demonstrate the differing characteristic scattering profile and easily distinguished onset energy. (c) Nonlinear least squares fit of the region between 400 eV and 700 eV to spectra corresponding to oxygen, platinum and carbon. The concentration map clearly shows the oxygen concentration in the edge of the particle yet absent in the other regions of the particle, consistent with the ADF signal.

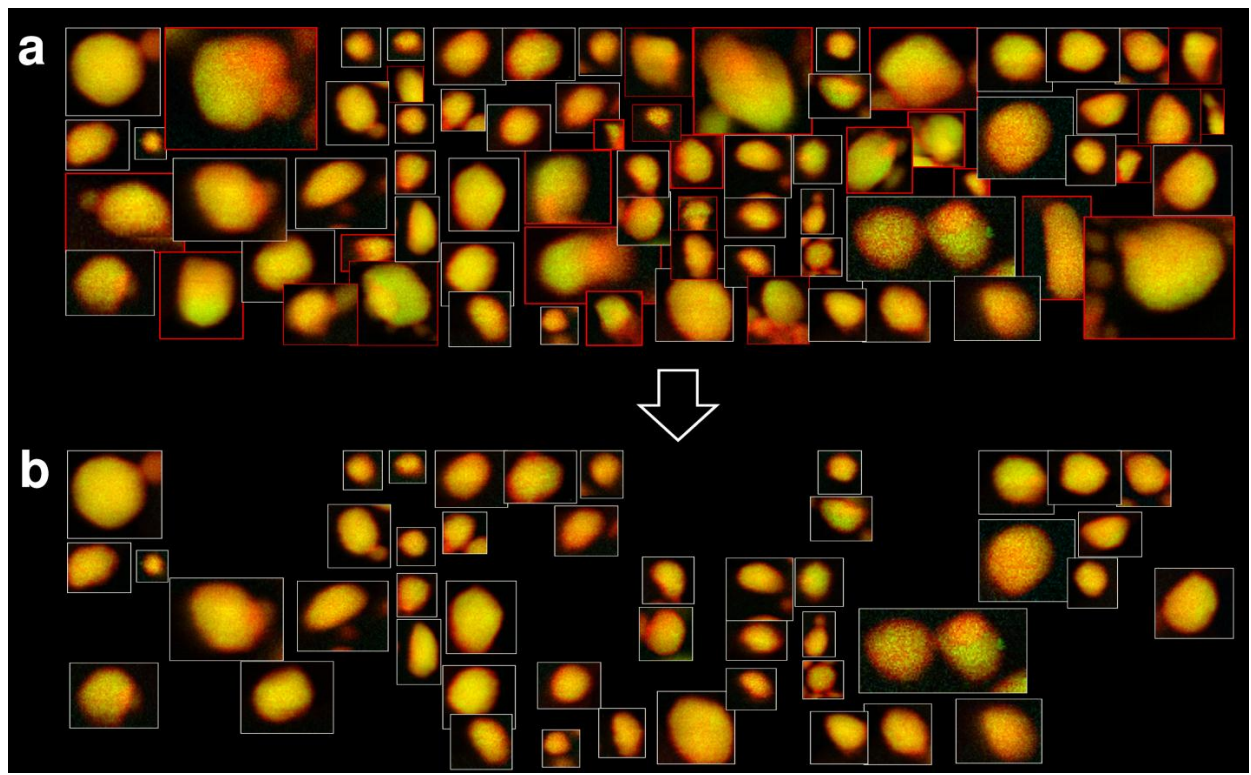

**Figure S15**. Spectroscopic images were recorded from a wide range of particle sizes in the as-received particle distribution as shown in (a). Because we used an automated thresholding method (see Figure S10) to extract the shell thickness for the as-received particles, we excluded the maps from 29 particles that could render failure or bias in the automated process. These include large particles that are sitting on a carbon support of variable thickness—with the change in the carbon thickness across the particle, the noise in the Co signal changes dramatically. (b) The remaining particles which were included in Fig. 4a and the analysis in Fig. 5 and Table 1. All particles are shown here in (a) to demonstrate that the profiles qualitatively match those of the remaining particles analyzed even if the quantitative analysis would be difficult.

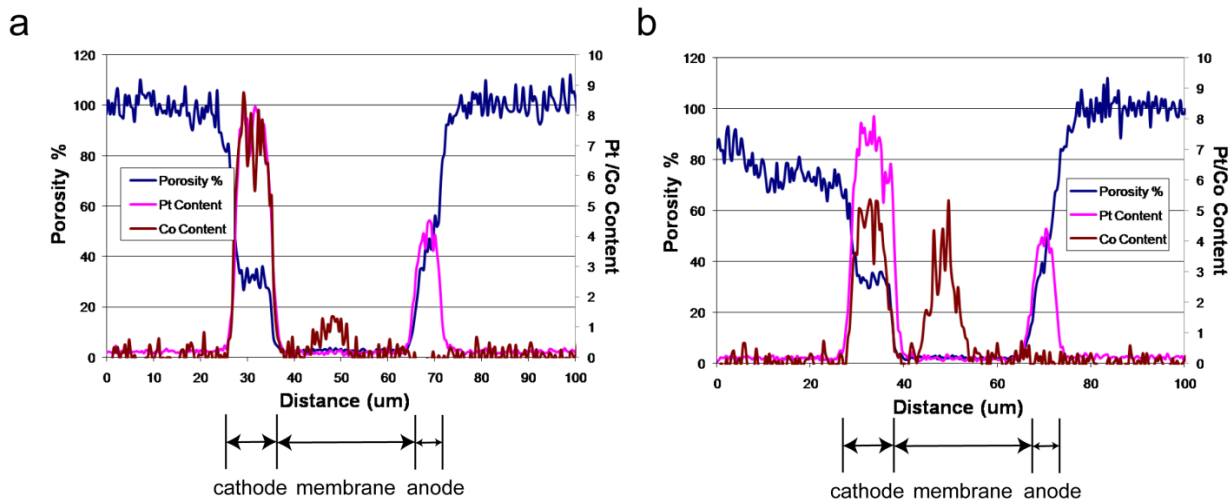

**Figure S16**, Electron microprobe analysis of the membrane electrode analysis prior to (a) and following 30,000 voltage cycles (b). There is a significant migration of Co from the cathode to the membrane after voltage cycling. (This analysis was performed by R. A. Waldo and I. Dutta at GM)